\documentclass[journal=nalefd, manuscript=letter, layout=twocolumn]{achemso} 
\usepackage{filecontents}
\usepackage{amssymb, physics, titlesec}
\usepackage[version=4]{mhchem}
\usepackage[separate-uncertainty = true]{siunitx}
\usepackage[hidelinks]{hyperref}
\usepackage{xcolor}
\usepackage[normalem]{ulem}

\makeatletter
\let\l@addto@macro\relax
\makeatother
\usepackage[fontsize=10pt]{scrextend}
\mciteErrorOnUnknownfalse

\titleformat{\subsection}[runin]{\bfseries}{}{}{}[]
\let\oldmaketitle\maketitle
\let\maketitle\relax

\makeatletter
\newcommand*{\addFileDependency}[1]{
\typeout{(#1)}
\@addtofilelist{#1}
\IfFileExists{#1}{}{\typeout{No file #1.}}
}\makeatother
\newcommand*{\myexternaldocument}[1]{%
\externaldocument{#1}%
\addFileDependency{#1.tex}%
\addFileDependency{#1.aux}%
}
\usepackage{xr}
\myexternaldocument{supp}

\title{Two Biexciton Types Coexisting in Coupled Quantum Dot Molecules}

 \author{Nadav Frenkel}
\affiliation{Department of Physics of Complex Systems, Weizmann Institute of Science, Rehovot 7610001, Israel}
\alsoaffiliation{Equal Contributor to This Work}

\author{Einav Scharf}
\affiliation{Institute of Chemistry and the Center for Nanoscience and Nanotechnology, The Hebrew University of Jerusalem, Jerusalem 91904, Israel}
\alsoaffiliation{Equal Contributor to This Work}

\author{Gur Lubin}
\affiliation{Department of Physics of Complex Systems, Weizmann Institute of Science, Rehovot 7610001, Israel}

\author{Adar Levi}
\affiliation{Institute of Chemistry and the Center for Nanoscience and Nanotechnology, The Hebrew University of Jerusalem, Jerusalem 91904, Israel}

\author{Yossef E. Panfil}
\affiliation{Institute of Chemistry and the Center for Nanoscience and Nanotechnology, The Hebrew University of Jerusalem, Jerusalem 91904, Israel}

\author{Yonatan Ossia}
\affiliation{Institute of Chemistry and the Center for Nanoscience and Nanotechnology, The Hebrew University of Jerusalem, Jerusalem 91904, Israel}

\author{Josep Planelles}
\affiliation{Departament de Quimica Fisica i Analitica, Universitat Jaume I, E-12080, Castello de la Plana, Spain}

\author{Juan I. Climente}
\affiliation{Departament de Quimica Fisica i Analitica, Universitat Jaume I, E-12080, Castello de la Plana, Spain}
\alsoaffiliation{Corresponding Author}
\email{climente@qfa.uji.es}

\author{Uri Banin}
\affiliation{Institute of Chemistry and the Center for Nanoscience and Nanotechnology, The Hebrew University of Jerusalem, Jerusalem 91904, Israel}
\alsoaffiliation{Corresponding Author}
\email{uri.banin@mail.huji.ac.il}

\author{Dan Oron}
\affiliation{Department of Molecular Chemistry and Materials Science, Weizmann Institute of Science, Rehovot 76100, Israel}
\alsoaffiliation{Corresponding Author}
\email{dan.oron@weizmann.ac.il}
 
\begin{document}

\twocolumn[
\begin{@twocolumnfalse}
\oldmaketitle

\begin{abstract}
Coupled colloidal quantum dot molecules are an emerging class of nanomaterials, introducing new degrees of freedom for designing quantum dot-based technologies. The properties of multiply excited states in these materials are crucial to their performance as quantum light emitters but cannot be fully resolved by existing spectroscopic techniques. Here we study the characteristics of biexcitonic species, which represent a rich landscape of different configurations, such as segregated and localized biexciton states. To this end, we introduce an extension of \textit{Heralded Spectroscopy} to resolve different biexciton species in the prototypical CdSe/CdS coupled quantum dot dimer system. We uncover the coexistence and interplay of two distinct biexciton species: A fast-decaying, strongly-interacting biexciton species, analogous to biexcitons in single quantum dots, and a long-lived, weakly-interacting species corresponding to two nearly-independent excitons separated to the two sides of the coupled quantum dot pair. The two biexciton types are consistent with numerical simulations, assigning the strongly-interacting species to two excitons localized at one side of the quantum dot molecule and the weakly-interacting species to excitons segregated to the two quantum dot molecule sides. This deeper understanding of multiply excited states in coupled quantum dot molecules can support the rational design of tunable single- or multiple-photon quantum emitters. 

    \end{abstract}

    \end{@twocolumnfalse}
]

Since the introduction of colloidal quantum dots (QDs) a few decades ago, their research is constantly developing, due to the intriguing quantum confinement effect that influences the electronic and optical properties as a function of the QD’s size and shape.\cite{GarciadeArquer2021,Talapin2010} QDs are impressively already widely implemented in commercial displays\cite{Panfil2018} and are of further relevance in additional applications including lasers,\cite{Park2021} light emitting diodes (LEDs),\cite{Won2019,Kim2020} single photon sources,\cite{Senellart2017} and photovoltaics.\cite{Kramer2011,Kagan2016} The extensive study in this field established synthetic means to allow for better control over the size, morphology, and surface chemistry of QDs of various semiconductor materials, enabling improved quantum yields (QY) and tunable emission and absorption spectra.\cite{Owen2017,Smith2010,Boles2016} In recent years, further research has been carried out to synthesize more complex nano-structures with two or more coupled emission centers, thus launching a new field of ``nano-chemistry".\cite{Bayer2001,Stinaff2006,Alivisatos1996,Battaglia2005,Deutsch2013} In particular, it was demonstrated that two QDs can be fused together \textit{via} a process of constrained oriented attachment, forming a coupled QD molecule (CQDM).\cite{Cui2019,Cui2021,Cui2021a,Koley2021} 

As QDs are often described as “artificial atoms” due to their discrete electronic states,\cite{Banin1999} CQDMs are in many senses analogous to artificial molecules,\cite{Alivisatos1996a} manifesting hybridization of the charge carrier wave functions. For the particular case of CdSe/CdS CQDMs, electron wave functions hybridize, whereas the hole wave function is localized to the cores due to the quasi-type II band alignment, the relatively large valence band offset between CdSe and CdS, and the heavier effective mass of the hole.\cite{Panfil2019} CQDMs exhibit optical and electronic properties which differ from their single QD building blocks as a result of the coupling.\cite{Cui2019} Notably, CQDMs’ spectrum is red-shifted and broader,\cite{Cui2021a,Koley2022,Verbitsky2022} the absorption cross-section is modified to be doubled at high energy and smeared out near the band gap,\cite{Panfil2019} their fluorescence decay lifetime is shorter, and their brightness is higher than their single QD constituents.\cite{Koley2022} The optical properties of the CQDMs depend on the width of the interfacial area between the two fused QDs, or ``neck'', serving as a potential barrier. The neck can be tuned chemically during the fusion process and was found to control the extent of the coupling and thus the electronic and optical properties.\cite{Cui2021a,Koley2022}
Moreover, the joining of two light emitting centers and the increase in the volume can stabilize both charged- and multi- electron--hole pairs (i.e., excitons), relative to such states in the respective single QDs, which are generally dimmed. The unique structure of the CQDMs can accommodate new types of multi-excitonic states and different relaxation pathways.\cite{Koley2022} In the simplest case of a biexciton (BX; two excitons occupying the same CQDM), the excitons can arrange in multiple spatial configurations within these nano-structures, whereas single QDs can only accommodate a single BX spatial configuration.\cite{Koley2022} 

 \begin{figure*}[!ht]
    \centering
    \includegraphics[width = \textwidth]{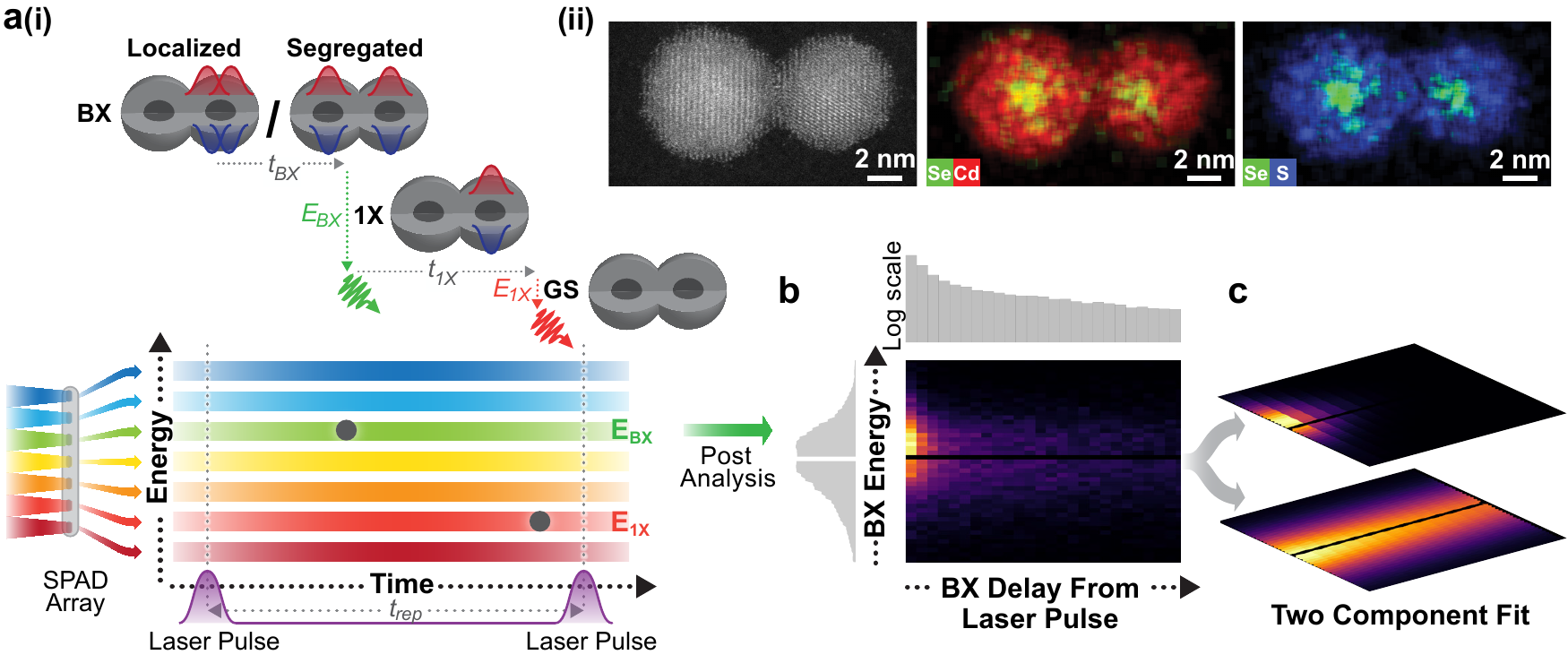}
     \caption{\textbf{Multiple BX States in CQDMs and the Heralded Spectroscopy Method. a)}(i) Top: Two photons are emitted sequentially by a radiative relaxation in a CQDM from a biexciton (BX) state of two possible spatial configurations, to the exciton (1X) state and eventually to the ground state (GS). Bottom: Scheme of the heralded spectroscopy method that uses photon correlations to resolve the arrival time and energy of the photon pairs. Only two-photon cascades that were detected following the same excitation pulse are registered as heralded events. (ii) High-angle annular dark-field scanning transmission electron microscopy (HAADF-STEM) image and energy-dispersive spectroscopy (EDS) images of a fused dimer. \textbf{b)} 2D spectrum-lifetime histogram of all the post-selected BX emissions from a 5-min measurement of a single CQDM. On top is the full vertical binning in logarithmic scale and to the left is the full horizontal binning of the 2D BX histogram, showcasing the BX decay lifetime and spectrum, respectively. \textbf{c)} The two-component fit of the BX population in (b), each component with an independent exponential decay in time and an independent Voigt profile distribution in energy. The black horizontal line in (b) and (c) is due to a `hot’ excluded pixel in the detector (see Methods section).}
    \label{fig:method}
\end{figure*}

Due to exciton--exciton interactions, the BX emission in many cases is spectrally shifted from the single exciton (1X) emission.\cite{Lubin2021HeraldedDots} In addition, multiple recombination pathways and non-radiative processes for BXs, such as the efficient Auger recombination, reduce the fluorescence decay lifetime of the BX, relative to that of the 1X.\cite{Park2014a} Therefore, a better understanding of BXs in nanocrystals is crucial for their incorporation in various applications, such as in lasing media, LEDs, and photovoltaics.  In CQDMs, this could help reveal some of their coupling properties towards more extensive control over their multi-excitonic characteristics. However, characterization of BX emission is challenging, as they generally cannot be spectrally separated at room temperature from the neutral and charged excitonic events, due to spectral diffusion and thermal broadening.\cite{Lubin2021HeraldedDots} Most of the previous work in this field utilized indirect methods to characterize the BX emission. The prevalent methods were power-dependent photoluminescence and transient absorption measurements, which exhibited a large variance in results.\cite{Ashner2019,Oron2006,Shulenberger2019,You2015,Sitt2009,Castaneda2016,Steinhoff2018,Oron2007} Recently, direct approaches to probe BX emission events at the single particle level were introduced, such as cascade or heralded spectroscopy.\cite{Lubin2021HeraldedDots,Vonk2021} These newly developed methods enable the energetic and temporal detection of sequential photons, thus eliminating the ambiguity associated with indirect methods. The heralded spectroscopy technique utilizes a \textit{spectroSPAD} system, which includes a single photon avalanche diode (SPAD) array at the output of a grating spectrometer.\cite{Lubin2021HeraldedDots} This system enables the post-selection of cascaded BX--1X events in a time-resolved-spectrally-resolved manner at room temperature. Therefore, it serves as an excellent tool for BX characterization in complex nano-structures such as CQDMs. Previous studies that utilized heralded spectroscopy, used temporal photon correlation between BX and 1X emissions to measure the BX shift ($\Delta_{BX} \equiv E_{1X}-E_{BX}$; the difference between the spectrum peaks of the 1X and of the BX emissions) at room temperature in single CdSe/CdS/ZnS quantum dots,\cite{Lubin2021HeraldedDots} and in CsPbBr$_3$ and CsPbI$_3$ perovskite nanocrystals (NCs).\cite{Lubin2021ResolvingSpectroscopy}

Herein, we explore the BX events in CQDMs and compare their properties to those of their constituent QDs, presenting an expansion of the powerful heralded spectroscopy methodology. Studying the prototypical system of CdSe/CdS core/shell CQDMs, we establish the coexistence of two BX species characterized by different lifetimes and 1X--1X interactions. Combining the experimental results with theoretical analysis, we attribute these to two BX spatial configurations. One where two holes are localized in the same QD (localized biexciton; LBX) and one where the two holes are segregated to the two constituent QDs (segregated biexciton; SBX), as illustrated at the top of \autoref{fig:method}a(i).  

\section*{RESULTS AND DISCUSSION}
\label{sec:results} 
The model system under study constitutes of CdSe/CdS CQDMs formed via the template approach introduced previously.\cite{Cui2019,Cui2021a} Briefly, CdSe/CdS core/shell QDs (radius of $1.35/\SI{2.1}{nm}$; electron microscopy characterization in \autoref{fig:method}a(ii) and \autoref{SIfig:TEM}), were bound to surface-functionalized silica spheres of $\SI{200}{nm}$ in diameter, followed by controlled coverage by an additional layer of silica, which blocks the unreacted silica binding sites and partially covers the QDs' surface, reducing the possibility to generate oligomers. Then, a molecular linker was added followed by the addition of a second batch of the same QDs, thereby attaching to the bound QDs, forming dimers on the template. Dimers are released via selective etching of the silica spheres by hydrofluoric acid and then undergo a fusion process at a moderate temperature. Size-selective separation is performed using the controlled addition of an anti-solvent, yielding a sample of the CQDMs.

In previous works that utilized heralded spectroscopy, extracting the BX emission spectrum was sufficient for a comprehensive BX characterization.\cite{Lubin2021HeraldedDots,Lubin2021ResolvingSpectroscopy} In the current case, the analysis is extended to resolve the BX population both spectrally and temporally, in order to account for the multiple BX species assumed to coexist in CQDMs. To explore BX states in CQDMs, cascaded emission events are directly probed at room temperature, extracting both temporal and spectral information simultaneously. The setup relies on exciting a single particle with a pulsed laser excitation, dispersing the emitted fluorescence by a grating spectrometer, and detecting the photons (temporally- and spectrally-resolved) with a SPAD array detector. Occurrences of photon-pair emission detected following the same excitation pulse are post-selected and treated as heralded events. Each photon within the post-selected photon pairs is time- and energy-tagged according to its time and pixel of detection. The high spectral and temporal resolutions (see Methods section) enable an unambiguous temporal separation between the two detections, attributing the first arriving photon to emission from the BX state and the second photon to emission from the 1X state (\autoref{fig:method}a(i) bottom). 

 \begin{figure*}[hbtp]
    \centering
    \includegraphics[width = \linewidth]{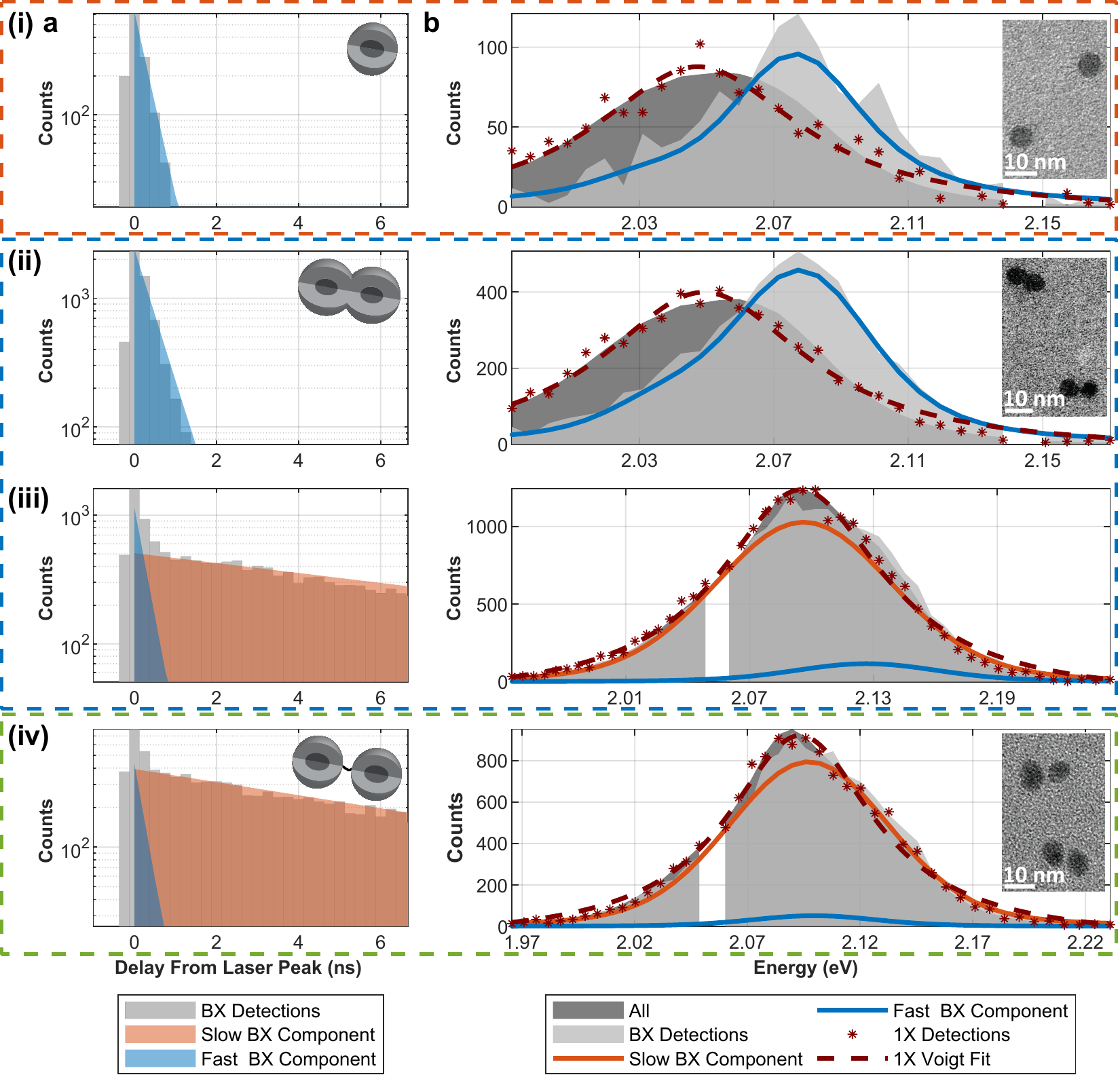}
     \caption{\textbf{2D Heralded Analysis of Single Particles.} The BX population from a 5-min measurement of (i) a monomer, a fused dimer with (ii) a high $g^{2}(0)$ contrast and (iii) a low $g^{(2)}(0)$ contrast, and (iv) a non-fused dimer. The particles feature a $g^{2}(0)$ contrast of approximately 0.09, 0.13, 0.37, and 0.45, respectively. Orange, blue, and green boxes distinguish between the different types of particles: monomers, fused dimers, and non-fused dimers, respectively. Schematics of the particle types are shown in the inset of (a) and transmission electron microscopy (TEM) images of the different particle types are shown in the inset of (b). The image of the fused dimers sample in panel (b) (ii) features two fused dimers that differ in the extent of fusion and filling of their interfacial area, the ``neck". \textbf{a)} The bright gray bars are the full vertical binning (FVB) of the 2D BX population histogram (as the one shown in \autoref{fig:method}b), showcasing the BX fluorescence decay lifetime. The blue and orange areas correspond to the FVB of the fast and the slow fitted BX components, respectively. A lifetime of $\SI{1}{ns}$ acts as a threshold between ``fast" and ``slow". \textbf{b)} The bright gray area is the full horizontal binning (FHB) of the 2D BX population histogram, showcasing the BX spectrum. The blue and orange lines correspond to the FHB of the fast and the slow fitted BX components, respectively. In red asterisks and red dashed line are the 1X spectrum and its fitted Voigt profile, respectively. In dark gray, the normalized spectrum of all detections from the measurement. The gap in the gray areas is due to a `hot' excluded pixel in the detector (see Methods section).}
    \label{fig:singleParticles}
\end{figure*}

Then, the BX population, in the form of a 2D spectrum-lifetime histogram (\autoref{fig:method}b), is fitted to the sum of two independent exponentially decaying components, using a least-squares solver (\autoref{fig:method}c): 

\begin{equation}
    f_{model} = a_1{\cdot}V_1(E)\frac{e^{-\frac{t}{\tau_1}}}{\tau_1} +a_2{\cdot}V_2(E)\frac{e^{-\frac{t}{\tau_2}}}{\tau_2}
\label{eq:model}
\end{equation}
  
Where $V_i(E)$ is a Voigt profile distribution in energy, i.e., over the detector's pixels, $\tau_i$ is the component's mono-exponential decay lifetime, and $a_i$ is a prefactor.

The results shown herein compare single NCs from two samples. One is of fused CQDMs, or ``fused dimers", and one is of ``non-fused dimers", where two QDs were linked together by the same template-based procedure described above, but not fused under moderate temperatures. Non-fused dimers remain connected by the molecular linker, but do not feature the continuous CdS lattice, i.e., the neck, between the QDs seen in \autoref{fig:method}a(ii). The dimer samples also contained single QDs, or ``monomers", that failed to attach to another QD (see Methods section and \autoref{SIfig:TEM} for further details). The monomers within the fused dimers sample were used as a reference for single QDs that underwent the same process. The photoluminescence signal from single particle measurements was used in several further analyses, allowing nanoparticle-type classification (\autoref{SIfig:blinking&FLID}), and collected under a single excitation power for all particle types (\autoref{SIfig:saturation}). The additional analyses included fluorescence intensity, intensity fluctuations, decay lifetimes, and the zero-delay normalized second-order correlation of photon arrival times ($g^{(2)}(0)$).\cite{Lubin2021HeraldedDots,Lubin2021ResolvingSpectroscopy} The first two supported nanoparticle-type classification for distinguishing monomers from dimers (following ref. \citenum{Koley2022}), while the $g^{(2)}(0)$ value was integral in revealing the nature of the NCs as quantum emitters, positioning them on the continuum between a single- and a multi-photon emitter. 

\autoref{fig:singleParticles} presents representative results of the 2D heralded analysis from 5-min measurements of (i) a monomer, (ii) a fused dimer with high $g^{(2)}(0)$ contrast, (iii) a fused dimer with low $g^{(2)}(0)$ contrast and (iv) a non-fused dimer. The left column (panels (a)) depicts the BX decay kinetics and the right column (panels (b)) shows the BX emission spectrum, both as bright gray areas. (i) and (ii) exhibited a single exponential BX decay, whereas (iii) and (iv) exhibited a bi-exponential BX decay. To graphically emphasize the difference between the NCs with a single exponential decay and the ones with a bi-exponential feature, the fitted components were labeled as ``fast" and ``slow", which refers to short and long BX decay lifetimes, respectively (blue and orange areas in panel (a), respectively). To distinguish between slow and fast decay patterns, a lifetime threshold of $\SI{1}{ns}$ was selected after preliminary results showed that monomers displayed only a sub-ns BX decay component (\autoref{SIfig:monomersBXLT}). The appearance of a ${>}\SI{1}{ns}$ BX lifetime in dimers (\autoref{SIfig:BXtwoTaus}) is therefore assumed to emanate from a BX species unavailable in monomers. Cases where the 2D fit exhibited two sub-ns components might be attributed to neutral and charged BXs,\cite{Xu2017} or simply to the additional degree of freedom in the fit. Consequently, in such cases, as in \autoref{fig:singleParticles}(i) and (ii), the two fast components are summed together and displayed as a single fast component, which represents well the observed decay. 

\autoref{fig:singleParticles}(i) presents a typical heralded spectroscopy characterization of a single monomer ($g^{(2)}(0){\approx}0.09$, see \autoref{SIfig:g2}). Panel (a) showcases a single sub-ns exponentially decaying fitted component (blue area; lifetime of $\tau{\approx}\SI{0.3}{ns}$), and panel (b) presents its spectrum (solid blue line). The BX shift of this component, i.e., the difference between the 1X peak (red dashed line) and the BX component's peak, is $\Delta_{BX}=-27{\pm}\SI{2}{meV}$ (all the error intervals in this work are given at 68$\%$ confidence levels). The negative BX shift (that is, a blue shift due to the 1X--1X repulsion) agrees well with a quasi-type II band-alignment regime, in which spilling out of the electrons wave functions to the shell reduces the overlap with the holes localized in the core, and hence the like-charges repulsion energies dominate over correlative attractions.\cite{Sitt2009} The normalized spectrum of all the detections from the 5-min measurement (including single photon events) is shown in dark gray and highly matches the 1X spectrum, indicating that the overall emission is dominated by 1X emission.  

\autoref{fig:singleParticles}(iv) presents a typical single non-fused dimer, which in contrast to the monomer in (i), displays two different components, with lifetimes of ${\sim}0.2$ and ${\sim}\SI{9}{ns}$ and different spectra (panels (a) and (b), respectively). The slow component (solid orange area) dominates the BX emission, with a relative contribution ($\frac{a_1}{\sum_{i=1}^{2} a_i}$) of ${\sim}97 \%$. The fast component (solid blue line in panel (b)) features $\Delta_{BX,fast}=-6{\pm}\SI{1}{meV}$. The emergence of a long-lived fitted BX component with a negligible shift ($\Delta_{BX,slow}=-4{\pm}\SI{1}{meV}$), is naturally associated with the multiple emission centers in this single non-fused dimer.  

In monomers, where the only possible BX spatial configuration is of two holes confined to one core (LBX), only a sub-ns strongly-interacting BX component is observed. Therefore, it is reasonable to attribute the reappearance of a similar fast-decaying BX component in the non-fused dimer, to LBX emission events. According to this reasoning, we assign the order-of-magnitude slower BX component to segregated BX (SBX) emission events. These assignments are validated by numerical simulations later in this section. Non-fused dimers consist of two nearly independent QDs, thus the SBX can be treated as two weakly-interacting 1Xs, as the two holes are separated into two different cores. Hence, the BX emission from such a state is expected to resemble the 1X emission in energy but with a shorter lifetime ($\tau_{1X}{\approx}\SI{18}{ns}$ in non-fused dimers), due to multiple recombination pathways and possible non-radiative competing processes, such as energy transfer or inter-core tunneling mechanisms.\cite{Cui2019} Moreover, in non-fused dimers, the long-lived BX is more dominant than the short-lived BX. This also supports their attribution to SBX and LBX states, respectively, as the non-radiative Auger decay dominates the LBX decay and reduces its QY. The near-zero $\Delta_{BX,slow}$ values are in agreement with the expected weak 1X--1X interaction. The small observed negative BX shift may be attributed to different quantum confinements of the constituent QDs. Bluer-emitting QDs feature shorter decay lifetimes (\autoref{SIfig:specVsLT}a), and hence the bluer-emitting QD of the non-fused dimer will more often emit first.

Fused dimers featured two distinct populations. Typical examples of each are seen in \autoref{fig:singleParticles}(ii) and (iii). Case (ii) resembles the monomer in (i) with its single sub-ns lifetime, $\Delta_{BX}=-25{\pm}\SI{1}{meV}$, and strong photon antibunching ($g^{(2)}(0){\approx}0.13$). Case (iii) has an emerging slow component (${\sim}\SI{11}{ns}$ lifetime), with a spectral offset ($\Delta_{BX,slow}=-1{\pm}\SI{1}{meV}$) resembling that of the non-fused dimer in (iv). This is accompanied by a weaker antibunching ($g^{(2)}(0){\approx}0.37$) compared to (i) and (ii). Case (iii) also shows a fast BX component that resembles the BX properties in (i) and (ii), featuring a lifetime of ${\sim}\SI{0.3}{ns}$ and $\Delta_{BX,fast}=-32{\pm}\SI{1}{meV}$. 

  \begin{figure}[t]
    \centering
    \includegraphics[width =\columnwidth]{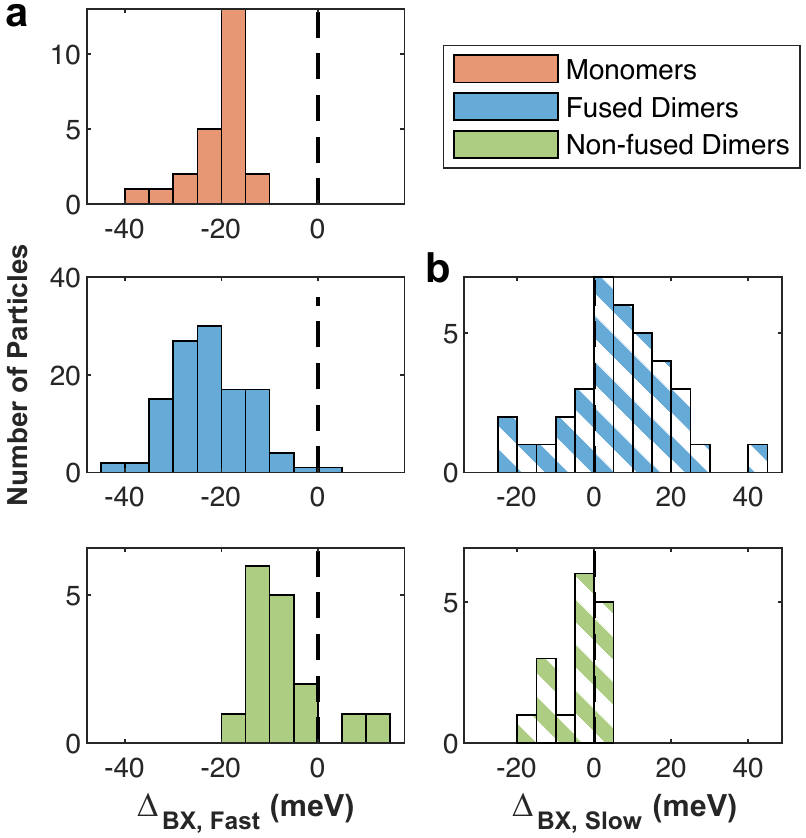}
     \captionof{figure}{\textbf{BX Shifts According to Particle Type.} BX shifts ($\Delta_{BX}$) of \textbf{(a)} the fast and \textbf{(b)} the slow fitted BX components of all the single particles, according to type. Monomers and ${\sim}68\%$ of the fused dimers did not exhibit a component with a lifetime of \SI{1}{ns} or higher and therefore do not appear in panel (b). Black dashed lines represent zero BX shift (equal energy of BX and 1X emissions).}
    \label{fig:BXHist}
\end{figure}

Moving to a statistical representation measured over numerous single particles, \autoref{fig:BXHist} displays the BX shifts of the fast and the slow components (panels (a) and (b), respectively) for all particles according to type. Monomers and ${\sim}68\%$ of the fused dimers presented only sub-ns BX components and therefore do not appear in panel (b). As shown in \autoref{fig:BXHist}a, monomers and fused dimers feature a similar $\Delta_{BX,fast}$ ($-21{\pm}\SI{6}{meV}$ and $-22{\pm}\SI{8}{meV}$, respectively). In non-fused dimers the fast BX shift is weaker ($\Delta_{BX,fast}=-7{\pm}\SI{7}{meV}$), due to stronger confinement effect. Unlike non-fused dimers, the monomers and fused dimers underwent ripening during the fusion process, which slightly thickened their shell, as apparent in their red-shifted emission (\autoref{SIfig:specVsLT}b). The lower volume of non-fused dimers increases the localization of electrons, which screens the Coulombic repulsion between the holes, reducing their fast BX shift, attributed to the LBX.\cite{Oron2007} Indeed, the different distributions of $\Delta_{BX,fast}$ for fused and non-fused dimers are in agreement with monomers that underwent the fusion process and monomers that did not, respectively (\autoref{SIfig:BXshiftsMonomers}). 

\autoref{fig:BXHist}b shows a slightly negative BX shift of $\Delta_{BX,slow}=-4{\pm}\SI{6}{meV}$ for non-fused dimers, consistent with the previously mentioned expectation of an averaged faster emission by the bluer-emitting QD within a dimer. Notably, 32$\%$ of the fused dimers also exhibited a slow component and showed $\Delta_{BX,slow}=8{\pm}\SI{15}{meV}$.

\begin{figure}[ht!]
    \centering
    \includegraphics[width = \columnwidth]{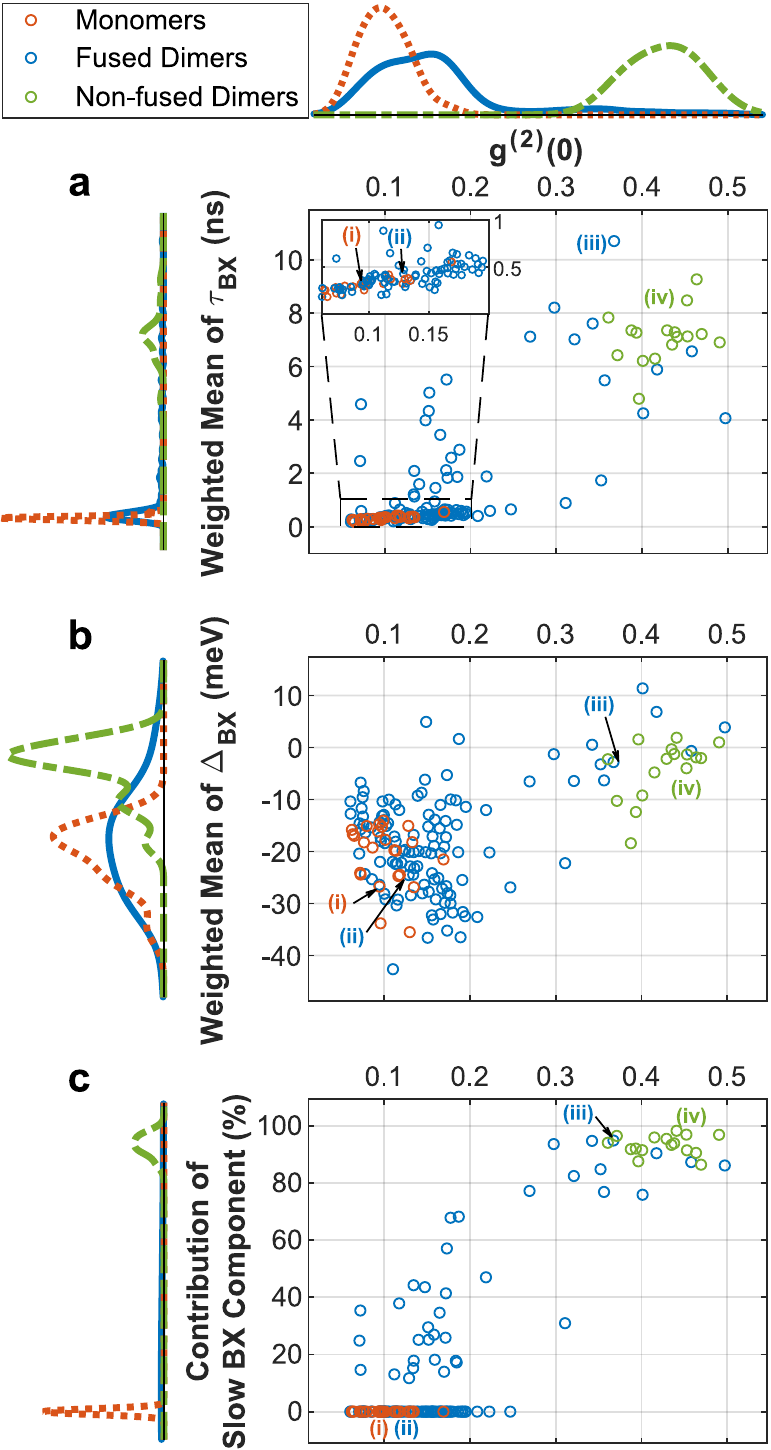}
     \caption{\textbf{BX Components as a Function of $g^{(2)}(0)$.} Weighted mean of \textbf{(a)} BX lifetimes and \textbf{(b)} BX shifts of the two fitted BX components of single particles and \textbf{(c)} the relative contribution of the ``slow"  BX component, as a function of $g^{(2)}(0)$, colored according to particle type. The particles shown in \autoref{fig:singleParticles} are marked with their corresponding number. Lines to the left and above the axes represent the marginal distributions as kernel density plots, with colors matching the particle type. In panel (c), the particles centered at 0 contribution are those that exhibited a sub-ns decay in both BX components.}
    \label{fig:BXAgg}
\end{figure}

\autoref{fig:BXAgg} shows the 2D heralded analysis of all particles as a function of the $g^{(2)}(0)$ contrast, indicating a single- or a multiple-photon emitter.\cite{Nair2011} This, with the exception of particles from the non-fused dimers sample that exhibited $g^{(2)}(0){>}0.55$, which were omitted from this work. This was to avoid the possible inclusion of oligomers or charged particles (see \autoref{SIsec:supporting analyses} and \autoref{SIfig:nonfusedG2}). The lifetimes of the two fitted BX components and their BX shifts (i.e., the difference between the spectrum peak of the 1X and the relevant BX component) are weighted according to the component’s relative contribution ($\frac{a_i}{\sum_{i=1}^{2} a_i}$). \autoref{fig:BXAgg}a shows that monomers display only a fast sub-ns BX dynamics, which agrees well with the LBX being the only available BX spatial configuration in monomers. The Auger recombination in such particles is highly efficient, leading to a high $g^{(2)}(0)$ contrast of $0.1{\pm}0.03$, classifying them as single photon emitters. 

The majority of the fused dimers exhibit high $g^{(2)}(0)$ contrasts (${<}0.2$) and a sub-ns BX lifetime, which we attribute to the LBX in those systems. Their BX shift distribution overlaps that of the monomers, as apparent in \autoref{fig:BXAgg}b. Consequently, these ‘monomer-like’ fused dimers can also be considered as single-photon emitters, yet with a larger absorption cross-section (see $\langle N \rangle$ estimation in Methods section) that increases the probability of multi-excitations.\cite{Panfil2019} Additionally, the larger volume at the neck region allows further electron delocalization in the LBX state, which reduces the efficiency of Auger recombination and slightly increases their emission intensity and BX yield (\autoref{SIfig:BXNum}). 

Together with increasing values of $g^{(2)}(0)$, the slow BX component emerges and eventually becomes the dominant one, as apparent in \autoref{fig:BXAgg}c and in the increase in the weighted mean of BX lifetimes and shifts (\autoref{fig:BXAgg}a,b). Notably, the BX shifts of each of the two BX states do not exhibit such correlation with $g^{(2)}(0)$ (\autoref{SIfig:BXshifts}). Accordingly, we assume that the observed trends in \autoref{fig:BXAgg} result from the varying ratio between the contribution of the segregated and localized BX states. 

Previous works showed that the neck thickness, which acts as a potential barrier, can control the extent of the electronic coupling, thus tuning the optical properties.\cite{Cui2021a,Koley2022} Generally, joining two emitting centers reduces photon antibunching due to the lower rate of the non-radiative Auger recombination of multiply excited states. However, by increasing the neck width, electron wave function delocalization partially retrieves the single photon source characteristics, increasing the $g^{(2)}(0)$ contrast.\cite{Koley2022} Therefore, we suggest that the trend of a decrease in photon antibunching is correlated with a decrease in the neck size. The position of the non-fused dimers at the edge of this trend (top right corner in \autoref{fig:BXAgg}a,b,c), with a negligible $\Delta_{BX}$ and a long BX lifetime, further validates the realization of the decrease in photon antibunching as a consequence of the decrease in the neck thickness. Non-fused dimers are separated by a linker, as mentioned earlier, and exhibit $g^{(2)}(0){\gtrsim}0.35$; therefore, they can be considered as two nearly-independent monomers. This sets the monomers and non-fused dimers as the extremes on the $g^{(2)}(0)$ scale, with the fused dimers distributed along it according to the extent of their neck filling.\cite{Koley2022} This trend demonstrates a unique property of CQDMs; by controlling their neck thickness, which acts as a synthetic tunable potential barrier, it is possible to continuously alter their behavior from a single-photon emitter to a two-photon emitter. 

\textbf{Quantum mechanical Simulations.} In order to establish a connection between the optical properties reported above and the morphological features of the CQDMs, we next carry out quantum mechanical simulations of the 1X and BX electronic structures in such systems. Our model is based on effective mass theory, which has successfully provided insight into the single exciton physics of CQDMs.\cite{Cui2019,Cui2021a,Panfil2019,Koley2022} Unlike in previous studies, however, we account for Coulomb interactions \textit{via} a Configuration Interaction (CI) procedure. Compared to the self-consistent method used in earlier works,\cite{Cui2019,Panfil2019} the CI method has the advantage of describing not only the ground state but also excited states. We shall see below that these can be relevant to understand the optical properties at room temperature.

To gain an understanding of the CQDMs' optical properties, we proceed in steps of increasing complexity. In the first step, we describe the BX shift in monomers through the ``BX binding energy" (i.e., the difference between twice the 1X \emph{ground} state energy and the BX \emph{ground} state energy). In the second step, we extend the analysis to BX and 1X excited states in CQDMs that are occupied at room temperature under thermal equilibrium. At this point, the comparison of the simulated spectrum with that observed in the experiments will allow us to infer information regarding the BX relaxation dynamics, and the existence of meta-stable excited states to explain the multi-exponential BX decay observed above.

 \begin{figure*}[!ht]
    \centering
    \includegraphics[width = \textwidth]{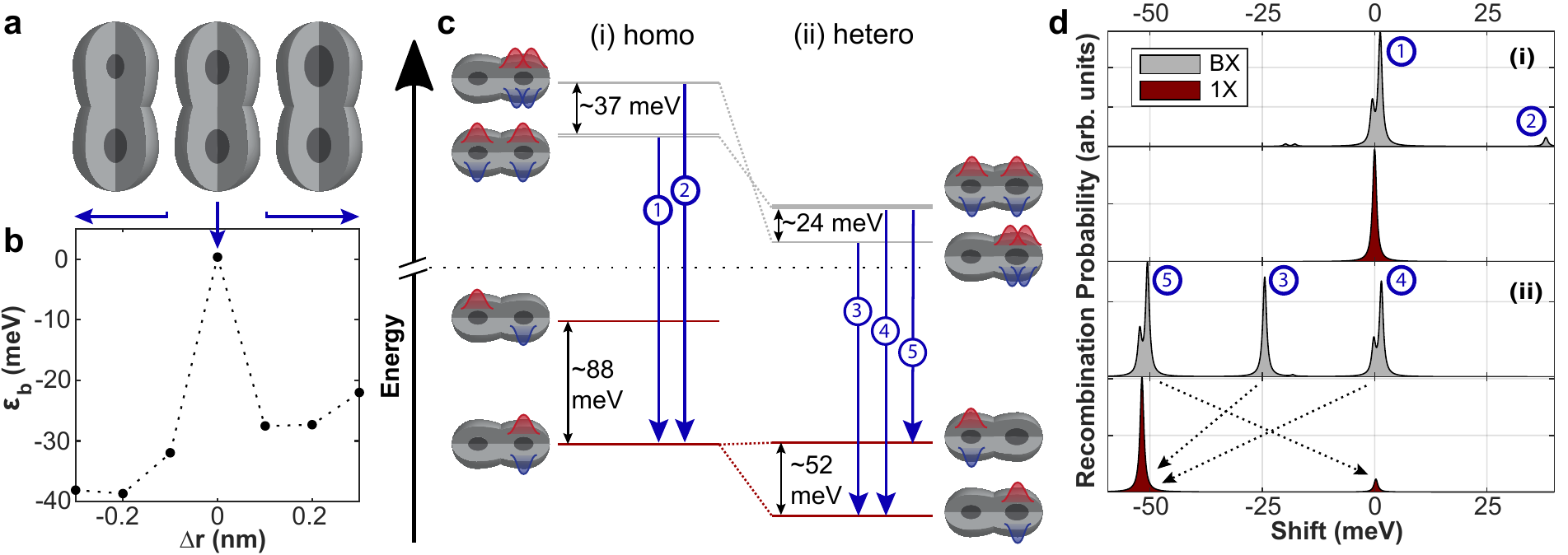}
     \caption{\textbf{Electronic Structures and Calculated Fluorescence Spectra of BX and 1X States in CQDMs. a)} Illustration of the homodimer (center) and heterodimers (left, right) under study. \textbf{b)} BX binding energy of the lowest-energy BX state as a function of the asymmetry between the sizes of the cores forming the CQDMs. Dots are calculated values and the dotted line is a guide to the eye. A small departure from the homodimer limit ($\Delta{r}{\approx}\SI{0.1}{nm}$) leads to a highly negative BX shift. \textbf{c)} Low-energy states of BXs and 1Xs in (i) homodimers ($\Delta{r}{=}0$) and (ii) heterodimers ($\Delta{r}{=}\SI{0.2}{nm}$). The blue arrows label the BX optical transitions, and the schematics illustrate the main charge carrier spatial configuration in the CI expansion. \textbf{d)} Simulated emission spectra of the BX (bright gray) and the 1X (red) in (i) homodimers and in (ii) heterodimers, as the ones shown in (c), at $T{=}\SI{300}{K}$. The reference energy (shift${=}\SI{0}{meV}$) is that of the 1X in the homodimers. The black dashed arrows indicate for transitions 3, 4 and 5 their respective resulting 1X state.}
    \label{fig:simulation}
\end{figure*}

Here, the BX binding energy is calculated as $\varepsilon_b{\equiv}2\varepsilon_{1X}-\varepsilon_{BX}$, where $\varepsilon_{1X}$ and $\varepsilon_{BX}$ are the ground state energies of single 1Xs and of BXs, respectively. Prior to the analysis of the complex dimer system, the monomer case was simulated. The monomers are approximated as spherical core/shell particles with a total (core+shell) diameter of $\SI{6.8}{nm}$. Negative binding energies, indicating repulsive 1X--1X interactions, in the same range as \autoref{fig:BXHist}a, are obtained for core radii between 1.25 and $\SI{1.55}{nm}$ (\autoref{SIfig:monomerEbSimulation}). In what follows, we consider QDs with a core radius of $\SI{1.35}{nm}$ that exhibit $\varepsilon_b{\approx}\SI{-35}{meV}$, which is a slightly stronger interaction than the mean BX shift for monomers in the experimental results. Next, we study the case of CQDMs (illustration in \autoref{fig:simulation}a). The CdSe cores are spherical, with radii $r_b$ and $r_t$ for the bottom and top cores, respectively. Each core has an ellipsoidal shell, with semi-axes $R_b$ and $R_t$ except in the coupling direction, where the semi-axes $n_b$ and $n_t$ define the neck filling.\cite{Panfil2019} The central CQDM in \autoref{fig:simulation}a illustrates a fused homodimer with $r_b{=}r_t{=}\SI{1.35}{nm}$, $R_b{=}R_t{=}\SI{3.4}{nm}$ (according to the size of the studied constituent QDs; see \autoref{SIfig:TEM}), and $n_b{=}n_t{=}\SI{7}{nm}$, which corresponds to a CQDM with a ‘rod-like' geometry (notice that a case of 
$n{=}R$ would imply no fusion at all).  

Because variations in the size of the QDs that constitute the CQDMs are likely to occur, we fix $r_b{=}\SI{1.35}{nm}$ and vary $r_t$. Thus, the left and right CQDMs in \autoref{fig:simulation}a schematically present small fluctuations in the size of the top core. \autoref{fig:simulation}b presents the BX binding energy of the lowest-energy BX state as a function of $\Delta{r}{=}r_t-r_b$. It follows from the figure that a precise homodimer ($\Delta{r}{=}0$) presents a weak binding energy ($\varepsilon_b{\approx}\SI{0}{meV}$), but as soon as heterogeneity in the core sizes comes into play, the BX binding energy switches to large negative values. Thus, for cores differing only in $\Delta{r}{=}{\pm}\SI{0.1}{nm}$, the CQDM already exhibits a BX binding energy of $\varepsilon_b{\approx}-\SI{30}{meV}$. It is then clear that core size fluctuations have a major influence on the BX energetics, whereas the neck size dispersion has a much weaker influence on it (\autoref{SIfig:simulatedNeckEffect}. Note that neck size can be expected to affect the BX \emph{dynamics} as discussed above). 

To understand the origin of this seemingly bimodal distribution of BX binding energies, in \autoref{fig:simulation}c we compare the electronic structure of a homodimer ($\Delta{r}{=}0$), and a heterodimer ($\Delta{r}{=}\SI{0.2}{nm}$). In the precise homodimer case (\autoref{fig:simulation}c(i)), the BX ground state is the SBX, with the LBX state blue shifted by ${\sim}\SI{37}{meV}$. The different stability stems from the nature of the 1X--1X interactions in each state. In the LBX state, these are repulsive intra-dot interactions, much like in the monomer, whereas in the SBX they are inter-dot interactions. Because 1X--1X interactions are dipole--dipole-like, they decay rapidly with distance. Inter-dot interactions are thus a minor effect, resulting in the SBX having about twice the energy of the 1X (i.e., $\varepsilon_b{\approx}\SI{0}{meV}$). The situation is, however, reversed in the heterodimer (\autoref{fig:simulation}c(ii)). When one of the cores is larger than the other, the LBX with both excitons in the large QD rapidly becomes the ground state. This is because the cores are in a strong quantum confinement regime, so relaxing the confinement easily overcomes the Coulomb repulsion between excitons. For this LBX state, $\varepsilon_b{\approx}\SI{-24}{meV}$. 

For a more direct comparison with the experiments, we next study how the electronic structures of homodimers and heterodimers translate into different optical spectra. \autoref{fig:simulation}d presents the calculated emission spectrum (assuming thermal equilibrium) of the 1X (red) and of the BX (bright gray) at room temperature for each type of CQDM, with the 1X emission in the homodimer acting as a reference point (shift${=}\SI{0}{meV}$). For the homodimer case (i), the 1X and BX spectra present a dominant peak at a similar energy. This is because at $\SI{300}{K}$, most BXs are in the SBX state, which has $\varepsilon_b{\approx}\SI{0}{meV}$ and relaxes to the direct exciton ground state (see arrow labeled as transition 1 in \autoref{fig:simulation}c). A small peak shows up at higher energies (transition 2), which corresponds to recombination from the LBX state, but its contribution is small as it originates from an excited state beyond thermal energy. 

In the heterodimer case (\autoref{fig:simulation}d(ii)), the 1X presents two peaks: a low-energy peak
corresponding to recombination in the larger QD, and a small peak at high-energy corresponding to
recombination in the smaller QD. The latter is small because of the scarce thermal occupation of the
excited 1X state (${\sim}\SI{52}{meV}$  above the ground state in \autoref{fig:simulation}c). The spectrum of the BX presents three relevant transitions. Transition 3 originates from the BX ground state, here the LBX. Its BX peak is blue-shifted from the main 1X transition by ${\sim}\SI{25}{meV}$. Transitions 4 and 5 correspond to recombination of an exciton in the smaller or in the larger QD, respectively. They arise from the SBX state, which can have some thermal occupation at room temperature if the core asymmetry is not large. Both transition 4 and 5 present a double peak fine structure (splitting of ${\sim}\SI{2}{meV}$). This feature is a consequence of the hybridization of the electron orbitals, forming bonding and anti-bonding molecular states, but it is not resolved in the experiments. It is also worth noting that both transition 4 and 5 present a similar intensity to 3,
despite the SBX being a few tens of meVs higher in energy than the LBX. This is because the SBX state is
highly degenerate (there are multiple ways to sort the two electrons and two holes in two QDs).
This property increases the chances of room temperature occupation for SBX up to a few tens of meV above the LBX ground state, such that an SBX
contribution can be expected except in CQDMs with severe core asymmetries. According to these calculations, the spectral width of the SBX emission (transitions 4 and 5) is expected to be greater than the one of the LBX emission (transition 3), which is supported by experimental results (\autoref{SIfig:BXFwhm}).  

We conclude from \autoref{fig:simulation}d that the BX optical emission of the homodimers is governed by the SBX, which has weak 1X--1X interactions, and hence emits at similar energies to the 1X. However, the BX emission energetics of heterodimers is richer, for it is governed by the LBX ground state with possible additional contributions from the SBX state. This occurs even in thermal equilibrium at room temperature and seems to account well for the observed energetic shifts of BXs for the case of heterodimers. Indeed, the heterodimer case is to be considered
when interpreting the experiments, since it is unlikely that the two cores forming a CQDM would be identical at the ångström level, which suffices to depart from the
homodimer limit according to our calculations.

The simulations shed light on the experimental results. They show that core heterogeneity must be assumed, resulting in LBX as the BX ground state, at least for the majority of fused dimers. This explains the fact that most of the fused dimers exhibited a dominant LBX emission, despite its strong quenching due to Auger recombination. The calculated  LBX binding energy of $\varepsilon_b{\approx}-\SI{25}{meV}$ also agrees with the values observed for the fast BX component shift in fused dimers, shown in \autoref{fig:BXHist}a. Nevertheless, considering the heterodimer limit, the observation of the  
${\sim}0$ BX shift in some fused dimers and in the majority of the non-fused dimers (\autoref{fig:BXAgg}b) is not fully explained by the calculations. Additionally, the simulations assume thermal equilibrium between the different BX states. This would result in a single BX lifetime averaged according to the Boltzmann distribution, contrasting the observed bi-exponential temporal decay in this aforementioned fraction of the fused and non-fused dimers. Therefore, to explain multiple BX radiative lifetimes, we must assume meta-stability for the different BX states. 
Moreover, since the neck size is shown to have a negligible effect on the calculated energetics, we posit that it does, however, have a significant impact on the BX relaxation dynamics, which is not captured by the static simulations.

We suggest that the BX emission greatly depends on the dynamics of BX relaxation to the lower-energy BX state, which can become much faster than the radiative BX recombination when the potential barrier is low (\autoref{SIfig:slowBXAmp}). Thus, assuming the case of heterodimers as mentioned earlier, the ‘hot’ generated SBX will relax with a high probability to form the lower-energy LBX in the larger QD of the pair. As the neck thickness decreases (corresponding to higher $g^{(2)}(0)$ in \autoref{fig:BXAgg}), relaxation from an SBX to an LBX becomes slower and thus less probable, as it competes with radiative processes. Because of a higher Auger rate in the LBX state, the SBX will become the dominant emitting BX in such a case, resulting in dimers with a multi-component BX emission characteristic. Indeed, the significant variable that changes along the decrease in photon antibunching and that agrees with its correlation with the mean BX shift in \autoref{fig:BXAgg}b , is the increasing ratio of SBX to LBX events (\autoref{fig:BXAgg}c). Consequently, the observed behavior of BXs in CQDMs is a result of the interplay between energetics, governed by size heterogeneity, and kinetics, governed by the potential barrier. 

 \section*{CONCLUSIONS}
\label{sec:conclusions}  

We resolve multiple biexciton species in the emission from coupled quantum dot molecules, introducing an extension to the powerful approach of heralded spectroscopy. Applying the technique to the prototypical CdSe/CdS coupled quantum dot dimers, single quantum dots, and non-fused dimers, revealed the coexistence and interplay of two biexciton species. Numerical simulations and experimental results attribute the fast-decaying, strongly-interacting biexciton species to localized biexcitons, where both holes are confined to the same CdSe core. The long-lived, weakly-interacting biexciton species is attributed to segregated biexcitons, where the two holes reside in the two CdSe cores. The relative contribution of each species correlates with the level of antibunching, ranging from single-photon emitters to two-photon emitters, and can be tuned continuously by controlling the width of the neck barrier between the constituent quantum dots. Finally, the numerical simulations also unveil the strong dependence of the energetics of the dimers' biexciton states on minute differences in the quantum dot core sizes, explaining the large percentage of dimers featuring a single quantum dot-like behavior. 

The unveiling of multiple biexciton species in coupled quantum dot molecules further demonstrates the potential of these materials as tunable and versatile quantum light emitters. Moreover, the extended heralded spectroscopy method applied here exemplifies the power and potential of this emerging spectroscopy technique to promote the understanding of nanocrystal photophysics and multiple-photon quantum emitters. 

\section*{METHODS}
\label{sec:methods}

\textbf{Synthesis of CQDMs and Sample Preparation.} The CdSe/CdS CQDMs were synthesized according to the protocol reported by Cui et al.,\cite{Cui2021a} using silica nanoparticles as a template. The template was used to link CdSe/CdS monomers through a thiol group. Additional SiO$_2$ was added to mask the exposed silica and immobilize the bound monomers. Introducing a second group of monomers after treating the first with a tetrathiol linker, formed dimer structures, attached by the linker. Then the silica nanoparticles were etched away via hydrofluoric acid treatment. Later, a ``strong” fusion process,\cite{Cui2021a} which includes prolonged heating, removed the linker and formed a uniform crystalline dimer. Size-selective separation excluded a large portion of monomers, resulting in a high dimer population. A dilute solution of NCs in 2.5$\%$ polymethyl methacrylate in toluene was spin-cast on a glass coverslip for the single-particle measurements. 

Three batches were used in this work (electron microscopy characterization in \autoref{SIfig:TEM}). The first is of monomers (referred to as “pristine monomers”) that did not undergo any further synthetic process and are used mainly for reference. The second is of fused dimers that underwent the fusion process with ``strong" fusion conditions ($\SI{240}{\degreeCelsius}$; 20 h; 5$\%$ ligands).\cite{Cui2021a} Since this procedure yields not just dimers but also some monomers (and some oligomers), the NCs from the fused dimers sample were classified according to their optical properties,\cite{Koley2022} to isolate the monomers in this sample (referred to as “monomers”. See inset in \autoref{fig:singleParticles}(i)b) from the dimers (see \autoref{SIsec:studiedQDs} for classification details). The last batch is of non-fused dimers, meaning pristine monomers that were linked and instead of undergoing the fusion process, were only heated for 1 h at $\SI{120}{\degreeCelsius}$ (see inset in \autoref{fig:singleParticles}(iv)b). This study displays results for single-particle measurements in which 400 BXs or more were detected, which amounted to 14 pristine monomers, 24 monomers out of the fused dimers sample, 116 fused dimers, and 16 non-fused dimers. 

\textbf{Optical Setup.} The SPAD array spectrometer is built around a commercial inverted microscope (Eclipse Ti-U, Nikon). An oil immersion objective (×100, 1.3 NA, Nikon) focuses light from a pulsed laser source (470 nm, 5 MHz, LDH-P-C-470B, PicoQuant) on a single particle (QD or CQDM) and collects the emitted photoluminescence. The emitted light is then filtered through a dichroic mirror (FF484-FDi02-t3, Semrock) and a long-pass filter (BLP01-473R, Semrock). The magnified image plane (×150) serves as the input for a Czerny-Turner spectrometer that consists of a 4-f system (AC254-300-A-ML and AC254-100-A-ML, Thorlabs) with a blazed grating (53-*-426R, Richardson) at the Fourier plane. At the output image plane of the spectrometer, a 512-pixel on-chip linear SPAD array is placed. Only fixed quarters of 64 pixels can participate simultaneously in the time-tagging measurement, which is done by an array of 64 time-to-digital converters (TDCs) implemented on a field programmable gate array (FPGA). The physical pixel pitch is $\SI{26.2}{um}$, which corresponds to a difference between neighboring pixels of ${\sim}\SI{1.7}{nm}$ in photon wavelength, or ${\sim}5-\SI{8}{meV}$ in energy. Of the single 64-pixel segment used in this work, the 34$^{th}$ pixel is a `hot' pixel and therefore omitted from all analyses. The instrument response function (IRF) of the system featured a ${\sim}\SI{190}{ps}$ full width at half maximum (FWHM). This response is a convolution of the excitation pulse temporal width and the timing jitter of the pixels. The pixels' dead time is ${\sim}\SI{15}{ns}$ and the average dark counts are ${\sim}41$ counts per second (CPS) per pixel. For further details on the experimental setup and analysis parameters see \autoref{SIsec:System and Analyses Parameters} and ref. \citenum{Lubin2021HeraldedDots}.

The laser illumination intensity was set to yield an average number of absorbed photons per particle per pulse ($\langle N \rangle$) of ${\sim}0.1$ for pristine monomers, calculated by saturation curves of the `on' state. Using the same analysis for fused dimers yielded 
$\langle N \rangle {\approx} 0.14$ (see \autoref{SIsec:System and Analyses Parameters} for further details). 

\textbf{Quantum Mechanical Simulations.} Calculations are carried within $k \cdot p$ theory framework. Non-interacting (single-particle) electron and hole states are calculated with the single-band Hamiltonians and material parameters of ref. \citenum{Panfil2019}, except for the relative dielectric constant inside the nano-structure which is here rounded to 10. In particular, we note that the conduction band offset is $\SI{0.1}{eV}$, which was found to provide good agreement with the experiments in earlier simulations of CQDMs.\cite{Panfil2019} Strain and self-energy corrections are disregarded for simplicity. Many-body eigenstates and eigenenergies are calculated within a full CI method, using \textit{CItool} codes.\cite{XY} Coulomb integrals for the CI matrix elements, including the enhancement coming from dielectric confinement, are calculated by solving the Poisson equation with Comsol Multiphysics 4.2. The CI basis set is formed by all possible combinations of the first 20 independent-electron and 20 independent-hole spin-orbitals. Charged exciton and biexciton configurations are then defined by all possible Hartree products between the few-electron and few-hole Slater determinants, consistent with spin and symmetry requirements. Optical spectra are calculated within the dipole approximation,\cite{XZ} assuming Lorentzian bands with a line-width of $\SI{0.5}{meV}$. Overall, the CI model is similar to that we have used to analyze other colloidal nano-structures, where the balance between carrier--carrier interactions is a key magnitude.\cite{YY,YZ} 

\section*{ASSOCIATED CONTENT}

\subsection*{Supporting Information.}

The supporting information (PDF) containing characterization of the studied quantum dots, description of the analyses parameters, and supporting experimental and theoretical analyses, is available free of charge.

\subsection*{Authors Information.}
\subsubsection*{Corresponding Authors}

{\quad}Juan I. Climente - Department de Quimica Fisica i Analitica, Universitat Jaume I, E-12080, Castello de la Plana, Spain; Email: climente@qfa.uji.es 

Uri Banin - Institute of Chemistry and the Center for Nanoscience and Nanotechnology, The Hebrew University of Jerusalem, Jerusalem 91904, Israel; Email: Uri.Banin@mail.huji.ac.il 

Dan Oron - Department of Molecular Chemistry and Materials Science, Weizmann Institute of Science, Rehovot 76100, Israel; Email: dan.oron@weizmann.ac.il

\subsubsection*{Authors}

{\quad}Nadav Frenkel - Department of Physics of Complex Systems, Weizmann Institute of Science, Rehovot 7610001, Israel

Einav Scharf - Institute of Chemistry and the Center for Nanoscience and Nanotechnology, The Hebrew University of Jerusalem, Jerusalem 91904, Israel

Gur Lubin - Department of Physics of Complex Systems, Weizmann Institute of Science, Rehovot 7610001, Israel

Adar Levi - Institute of Chemistry and the Center for Nanoscience and Nanotechnology, The Hebrew University of Jerusalem, Jerusalem 91904, Israel

Yossef E. Panfil\textsuperscript{\#} - Institute of Chemistry and the Center for Nanoscience and Nanotechnology, The Hebrew University of Jerusalem, Jerusalem 91904, Israel

Yonatan Ossia - Institute of Chemistry and the Center for Nanoscience and Nanotechnology, The Hebrew University of Jerusalem, Jerusalem 91904, Israel

Josep Planelles - Department de Quimica Fisica i Analitica, Universitat Jaume I, E-12080, Castello de la Plana, Spain

\textsuperscript{\#}\textit{Present Address}: 200 S. 33\textsuperscript{rd} St. 201 Moore Building, Philadelphia, PA 19104, USA

\textit{Author Contributions:
}
N.F. and E.S. contributed equally to this work.

\subsection*{Notes.}
The Authors declare no competing financial interest.

\section*{ACKNOWLEDGEMENTS}
\label{sec:acknowledgements}  
U.B. and D.O. acknowledge the support of the Israel Science Foundation (ISF), and the Directorate for Defense Research and Development (DDR\&D), grant No. 3415/21.
J.I.C. and J.P. acknowledge support from UJI project B-2021-06. E.S., A.L., Y.E.P., and Y.O. acknowledge support from the Hebrew
University Center for Nanoscience and Nanotechnology.

\bibliography{main.bib}

\makeatletter\@input{xx.tex}\makeatother
\end{document}


\maketitle
\centerline{Email:  climente@qfa.uji.es ; uri.banin@mail.huji.ac.il ; dan.oron@weizmann.ac.il}

\begin{abstract}
    This supporting information describes in greater detail the synthesis, data analysis, and system parameters used in this work, as well as additional analyses supporting the information given in the main text. 
Sections are brought in the order of their reference in the main text.
    \end{abstract}
\begin{multicols}{2}

\section{The Studied Coupled Quantum Dot Molecule Samples}
\label{SIsec:studiedQDs}

\textbf{Synthetic Procedure.} The synthetic procedure follows ref. \citenum{Cui2021a} and is described briefly in the Methods section in the main text.  

\begin{figure}[H]
    \centering
    \includegraphics[width=\columnwidth]{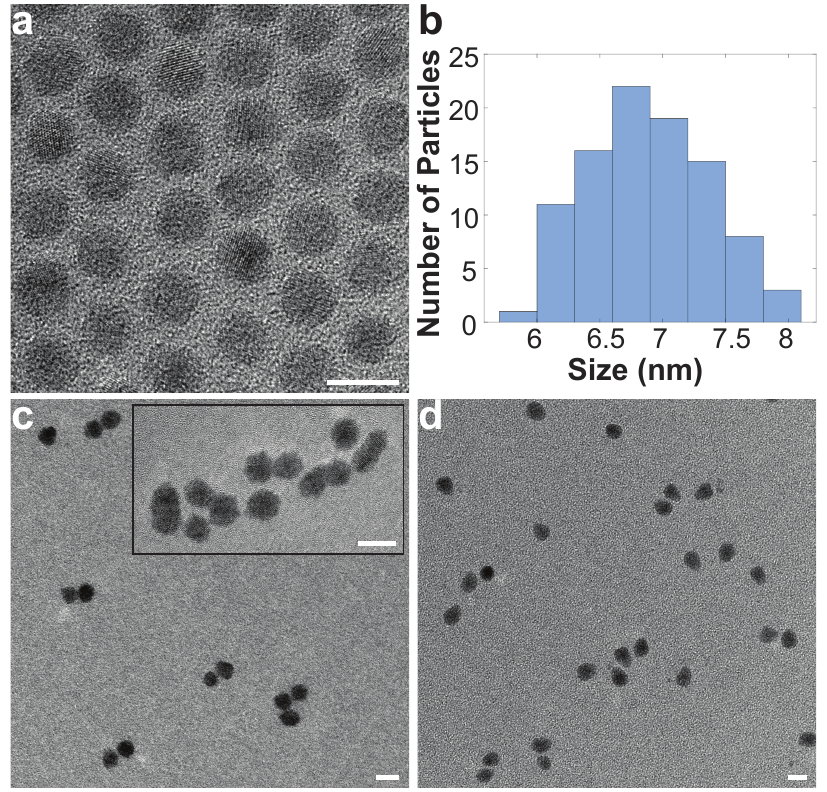}
    \caption{\textbf{Electron Microscopy Characterization of the Studied Samples. a)} High-resolution transmission electron microscopy (HR-TEM) image of “pristine monomers” (single QDs that were not exposed to fusion conditions). \textbf{b)} Size distribution of 95 pristine monomers with a mean diameter of $6.9\pm$\SI{0.5}{nm}. \textbf{c)} TEM image of fused dimers. HR-TEM image in the inset showcases different extents of filling of the connecting area between the fused monomers (the neck). \textbf{d)} TEM image of non-fused dimers. The dimer samples include monomers and multimers. All scale bars are \SI{10}{nm}.}
    \label{SIfig:TEM}
\end{figure}

\textbf{Particle Type Classification.} The size-selective precipitation at the end of the synthetic dimer formation separates monomers, dimers, and multimers. Yet, the separation is not full, and monomers and oligomers are still found in the dimer samples (see \autoref{SIfig:TEM}). According to transmission electron microscope (TEM) images, in the fused dimers sample ${\sim}50\%$ of the particles were dimers, and in the non-fused dimers sample ${\sim}25\%$ of the particles were dimers. Consequently, we cannot avoid single-particle measurements of all the species in these samples. Therefore, a classification procedure is required to distinguish between the single particle types. Here, we applied the classification process reported by Koley et al. and adapted it as described below.\cite{Koley2022} We note that the strength of the `spectroSPAD' as a comprehensive spectroscopy tool is demonstrated by the extraction of all the spectroscopic insights described below by post-processing of the same 5-min single-particle raw data collected for the heralded analysis described in the main text. Upon classifying the type of the measured single particles, it is apparent that the percentage of measured dimers exceeded their occurrence in the samples, as  ${\sim}83\%$ of the measured particles in the fused dimers samples were dimers. At least  ${\sim}35\%$ of the measured particles in the non-fused dimers sample were dimers (see \autoref{SIsec:supporting analyses} for further insight on the non-fused dimers as nearly uncorrelated single photon sources). We attribute these statistics to selection bias, selecting brighter spots in the sample (thus avoiding most monomers) and avoiding spatially extended spots (suspected to be aggregates) during the single-particle measurements. This might have increased the fraction of the measured dimers compared with the unbiased statistics collected by electron microscopy. 

The joining of two emitting centers and the different structure of dimers are manifested in different optical properties. Those differences help in distinguishing between monomers and dimers.\cite{Koley2022} Here we present additional analyses performed to allow particle type classification. 

\begin{figure}[H]
    \centering
     \begin{subfigure}[b]{0.5\textwidth}
         \centering
    \includegraphics[width=\columnwidth]{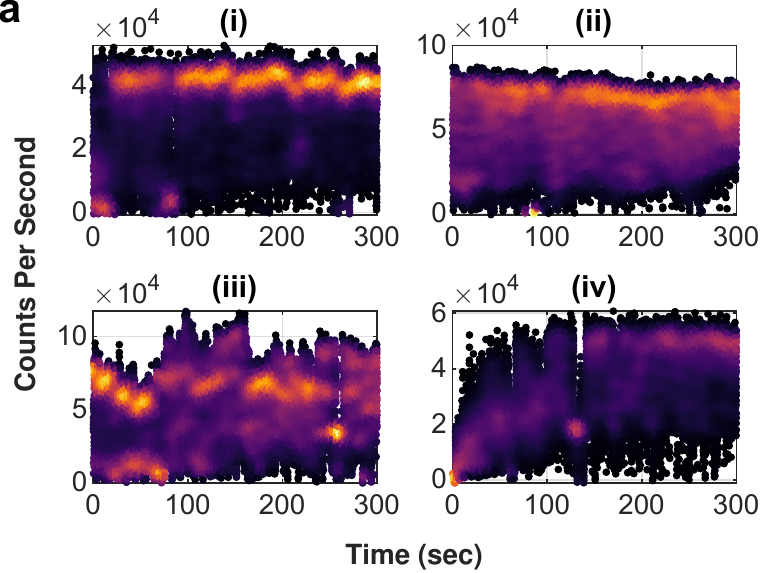}
        \end{subfigure}
        \hfill
     \begin{subfigure}[b]{0.5\textwidth}
         \centering
    \includegraphics[width=\columnwidth]{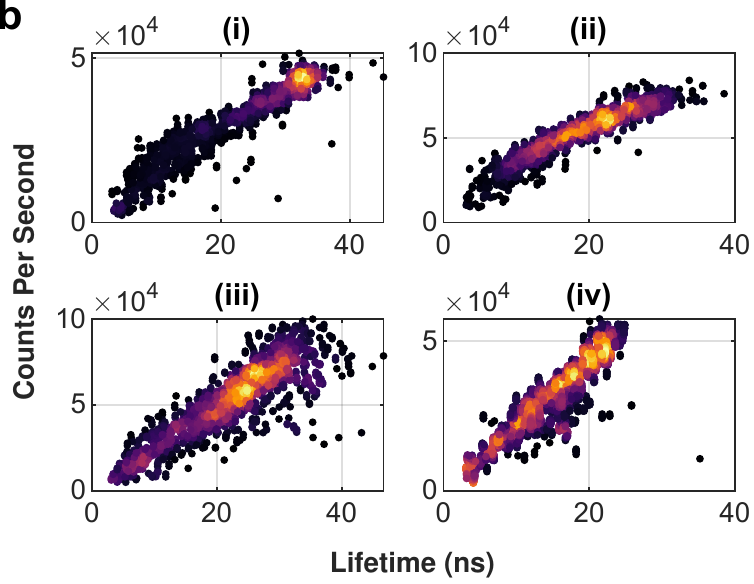}
        \end{subfigure}
        \caption{\textbf{Typical Optical Properties of Single Particles.} The single particles from \autoref{fig:singleParticles} in the main text are shown: (i) A monomer, a fused dimer with a (ii) high and a (iii) low \(g^{(2)}(0)\) contrast, and a (iv) non-fused dimer. \textbf{a)} Fluorescence intensity fluctuation time-trace. All detections from a 5-min measurement are binned into 10 ms bins according to the global detection time. The counts per second (CPS) for each bin are then calculated. \textbf{b)} fluorescence-lifetime-intensity distribution (FLID). Each dot represents a 50 ms time-bin within the 5-min measurement. The color indicates data-point density, where brighter areas correspond to a denser population.}
    \label{SIfig:blinking&FLID}
\end{figure}

Examples of these are shown in \autoref{SIfig:blinking&FLID} for the single particles from \autoref{fig:singleParticles} in the main text, where (i) is a monomer, (ii) and (iii) are fused dimers with high and low $g^{(2)}(0)$ contrasts, respectively, and (iv) is a non-fused dimer. \autoref{SIfig:blinking&FLID}a displays the intensity fluctuations (`blinking') of the single particles and \autoref{SIfig:blinking&FLID}b presents the fluorescence intensity-lifetime distribution (FLID). In \autoref{SIfig:blinking&FLID}b, detections are binned into $\SI{50}{ms}$ time bins. Each bin is assigned with an intensity value, by summing over all detections in that bin, and with an average lifetime, estimated as the temporal delay from the laser where the detections population drops by $\frac{1}{e}$. In the monomer, there is a clear ‘on’ state with an emission rate of ${\sim}4\cdot10^4$ counts per second (CPS), as apparent in \autoref{SIfig:blinking&FLID}a,b (i). Most photons are emitted from the `on' state that features a relatively high count rate and a long lifetime (\autoref{SIfig:blinking&FLID}b (i)). In dimers, the top count rate is higher than the ${\sim}4\cdot10^4$ CPS of the monomer (\autoref{SIfig:blinking&FLID}a,b (ii), (iii), and (iv)), which allows their identification. \autoref{SIfig:blinking&FLID}b (ii), (iii), and (iv) do not feature a well-defined `on' state, and emission is probable at different count rates. The lifetime of the frequent emitting states is shorter as well.

All of the observed differences can be associated with the larger absorption cross-section of the dimers and their larger volume compared to monomers. In monomers, Auger decay is efficient. Hence emission from charged and multi-excited states is dimmed. Most of the detections are emitted from the neutral exciton state with a high count rate and a long lifetime. In dimers, the higher count rate is attributed to the nearly two-fold absorption cross-section (see \autoref{SIsec:System and Analyses Parameters}).\cite{Panfil2019} The high volume of the dimers decreases the Auger rate, which increases the contribution of charged and multi-excited states. Emission from charged states, in particular, was found to be significant in dimers.\cite{Koley2022} Accordingly, in dimers, the peak of most detections is shifted toward intermediate count rates with a shorter lifetime (\autoref{SIfig:blinking&FLID}b (ii), (iii), and (iv)). The fused dimers in panels (ii) and (iii) also vary in some of their optical properties. For example, the fused dimer in panel (iii) exhibits stronger intensity fluctuations than the fused dimer in panel (ii). We later show that another differing feature between dimers is the $g^{(2)}(0)$ contrast (see \autoref{SIfig:g2}). These different properties were previously explained by variations in the potential barrier, governed by the neck thickness.\cite{Koley2022}  

The collective overview of the optical properties in \autoref{SIfig:blinking&FLID} helps to distinguish between monomers and dimers. The identification of multimers is done according to the intensity and \(g^{(2)}(0)\) contrast (see \autoref{SIsec:supporting analyses}).

\section{System and Analyses Parameters}
\label{SIsec:System and Analyses Parameters}
This section describes the measurement and analysis parameters and details some of the progress in the SPAD detector since previous accounts.\cite{Lubin2021HeraldedDots,Lubin2021ResolvingSpectroscopy}

\begin{figure}[H]
    \centering
    \includegraphics[width=\columnwidth]{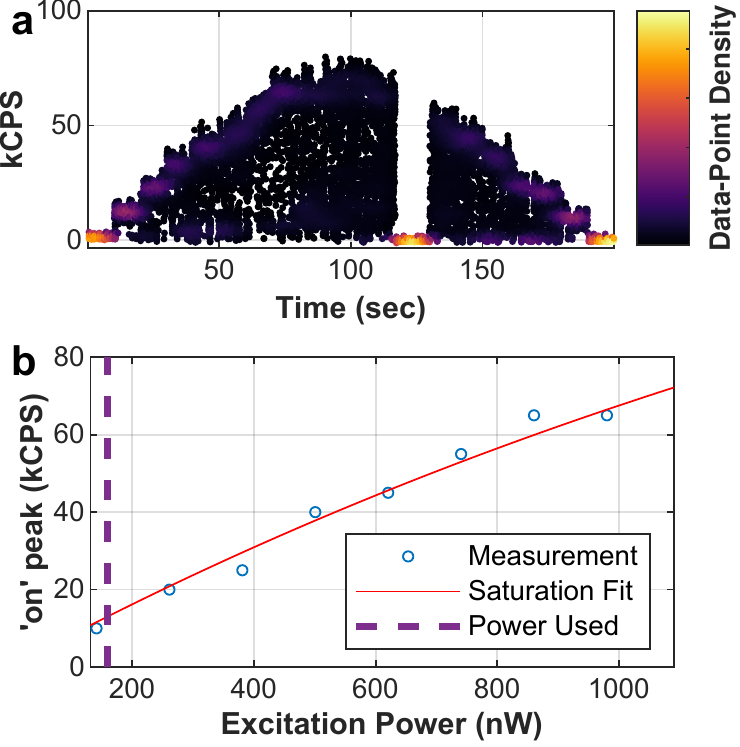}
    \caption{\textbf{Saturation Experiment of a Pristine Monomer. a)} Intensity fluctuation trace (shown as kilo-counts per second, i.e. kCPS) as a function of time for a single pristine monomer under varying illumination powers. The laser intensity increases in 10 seconds steps and then decreases back after reaching the maximum power. \textbf{b)} In blue circles, the intensity of the `on’ peaks at each illumination power. The red line indicates the saturation curve fit. The dashed purple line indicates the power used in the experiments of this study (160 nW). For this particle \(\langle N \rangle{=}0.08\pm0.07\) (68$\%$ confidence interval).}
    \label{SIfig:saturation}
\end{figure}

\textbf{\(\langle N \rangle\) Estimation.} To assess the saturation intensity, we follow the procedure in the Supporting Information of ref. \citenum{Lubin2021HeraldedDots}. Single pristine monomers were illuminated in varying intensities, increasing every 10 seconds up to some maximal value, and then decreasing following the same steps (see \autoref{SIfig:saturation}a). In order to assess the emission saturation, we plot the intensity histogram for each excitation power and identify the ‘on’ state peak. We use these data points to fit a saturation curve model (see \autoref{SIfig:saturation}b):\cite{Teitelboim2015}
\begin{equation}
    P = A(1-e^{I/I_{sat}})
\end{equation}

P is the `on’ state peak and I is the excitation power. The fitted parameters are $I_{sat}$, which is the saturation power, and A, the asymptotic ‘on’ state peak. We then estimate the average number of absorbed photons per excitation pulse as \(\langle N \rangle = \frac{I_{used}}{I_{sat}}\), where $I_{used}$ is the laser intensity used in the experiment (dashed purple line in \autoref{SIfig:saturation}b).

This model assumes negligible contribution by multi-excitation recombinations, which is validated by $g^{(2)}(0){\approx}0.09$ for pristine monomers under the illumination power in this study. For pristine monomers, we obtain \(\langle N \rangle{=}0.1\pm0.07\). The same procedure was done for single fused dimers and resulted in \(\langle N \rangle{=}0.14\pm0.11\). Dimers are the product of joining two monomers. Thus we expect a two-fold absorption cross-section. Yet, as discussed above, emission from charged states is significant, whereas emission from the ‘on’ state is less frequent. This results in an underestimation of the intensity of the ‘on’ state in each excitation power, which reduces the calculated \(\langle N \rangle\). Still, we can set an upper limit of \(\langle N \rangle{\approx}0.2\) for dimers, as they consist of two monomers.

\textbf{SPAD Detector.} The main difference between the experimental setup used in this work and the one depicted in refs. \citenum{Lubin2021HeraldedDots,Lubin2021ResolvingSpectroscopy} is an updated pixel wiring configuration of the detector by the manufacturer. Neighboring pixels are now active during the measurement, compared with an every-other-pixel configuration in the previous version. The higher fill factor leads to a two-fold enhancement of the single-photon detection probability, which translates to a dramatic four-fold enhancement in the photon-pair detection probability. The use of neighboring pixels also increases the probability of inter-pixel optical crosstalk (a detailed description of its characterization could be found in ref. \citenum{Lubin2019QuantumArrays}), which required closer attention to the corrections made. The crosstalk and dark counts contribution was calculated for each spectral-temporal bin of the 1X and BX 2D histograms and subtracted from the raw signal. 

\textbf{Analyses Parameters.} Sequential photon emissions that were both detected after the same laser excitation pulse were registered as heralded events, as depicted in the main text. This, providing that they met the following temporal and pixel constraints. The first detected photon of the pair (BX) was constrained between $-0.5$ and \SI{20}{ns} delay from the laser pulse peak. The lower gate is to accommodate for the instrument response function (IRF) of the system (some detections will seem to arrive before the excitation pulse due to detector jitter and excitation pulse width). The second photon of the pair (1X) was gated between 0.5 to \SI{60}{ns} delay from the first photon. The non-zero lower gate serves to exclude events where the detection order is not clear, and diminish the contribution of crosstalk events (both feature temporal response corresponding to the system IRF). The upper bounds for both detections (BX and 1X) are longer than their respective lifetimes but significantly shorter than the laser pulse period (\SI{200}{ns}). This was chosen to lower signal loss while maintaining low dark counts contributions and ensuring both photons originated from the same excitation pulse. In addition, because of the detection dead time, mentioned in the Methods section in the main text, sequential detections in the same pixel could only occur if the photons are \SI{15}{ns} apart or longer. Therefore, photon pairs that were detected at the same pixel were excluded entirely to prevent bias in favor of longer-lived photon cascades. After sifting the raw data for cascaded BX--1X events, statistical corrections for dark counts and crosstalk were applied to subtract false detections, following the scheme outlined in refs. \citenum{Lubin2021HeraldedDots,Lubin2021ResolvingSpectroscopy}. 

\begin{figure}[H]
    \centering
    \includegraphics[width=\columnwidth]{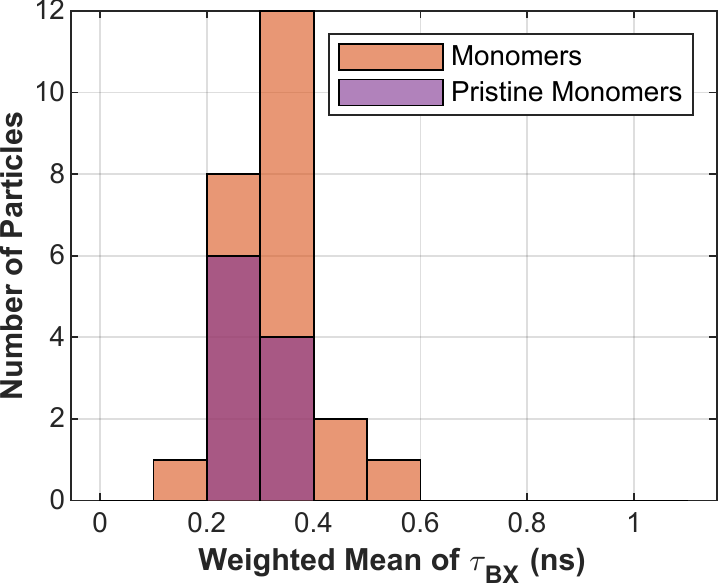}
    \caption{\textbf{BX Lifetime of Monomers and Pristine Monomers.} Weighted mean of BX lifetimes of the two BX components for monomers and pristine monomers.}
    \label{SIfig:monomersBXLT}
\end{figure}

\textbf{Biexciton Components Distinction.} Biexciton (BX) events (i.e., the first detected photon of each post-selected photon cascade) were fitted to two independent exponentially decaying components, as mentioned in the main text. The distinction of two BX sub-populations, one slowly- and one fast-decaying, was based on preliminary results for monomers and pristine monomers (see \autoref{SIfig:monomersBXLT}). These showed a weighted mean BX lifetime (calculation mentioned in the main text) of the two BX components no greater than \SI{0.6}{ns}. Therefore, a threshold of \SI{1}{ns} ns lifetime was chosen to distinguish between “fast” and “slow” BX components. 

\begin{figure}[H]
    \centering
    \includegraphics[width=\columnwidth]{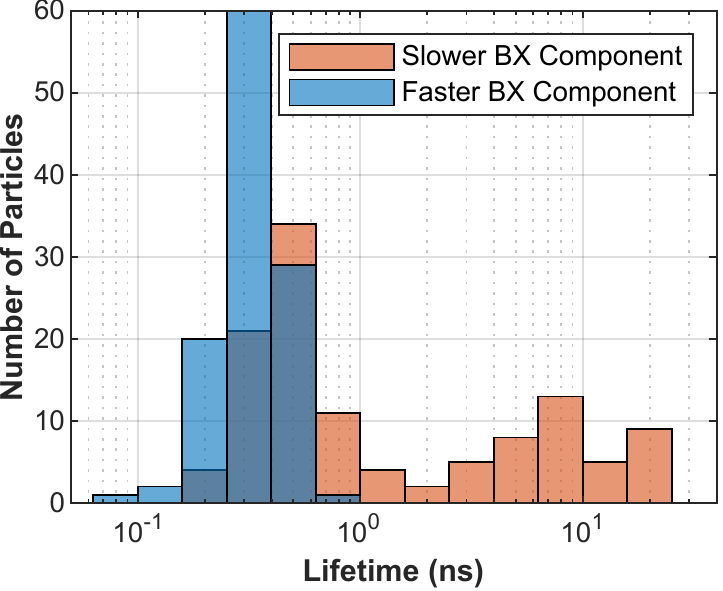}
    \caption{\textbf{Decay Lifetime of BX Components in Fused Dimers.} The lifetime distribution of the two BX components in all fused dimers, without categorizing them to ``slow" and ``fast" components as shown in other figures. In orange, the component with the longer lifetime out of the two and in blue, the one with the shorter lifetime.}
    \label{SIfig:BXtwoTaus}
\end{figure}

This distinction was made assuming that the emergence of a slow component in dimers would be due to a physical process not available in monomers. Indeed, \autoref{SIfig:BXtwoTaus} shows two distinct populations of fused dimers. All fused dimers have at least one BX component with a sub-ns lifetime (blue area). For most of them, the second BX component also has a sub-ns lifetime (orange area overlapping the blue area), while for the minority, the second component has a lifetime of $\SI{1}{ns}$ or above (orange area not overlapping the blue area).
 
\section{Supporting Analyses}
\label{SIsec:supporting analyses}
This section includes further analyses performed on the single-particle level. It describes the calculation for the zero-delay normalized second-order correlation of photon arrival times (\(g^{(2)}(0)\)) and presents it for the single particles in \autoref{fig:singleParticles} in the main text, as an example. Then, it follows with further aggregate analyses that support the information in the main text. All supporting analyses were performed on the same raw data collected for the heralded spectroscopy and used in the main text.

\textbf{$g^{(2)}(0)$ Calculation.} The $g^{(2)}(0)$ was calculated and corrected for errors emanating from dark counts and inter-pixel crosstalk according to the protocol detailed in ref. \citenum{Lubin2019QuantumArrays}. Briefly, each combination of SPAD array pixels pair is treated as the arms of a Hanbury Brown and Twiss photon correlation setup. Pairs of photon detections are counted according to the delay between them ($\tau$) and binned to ${\sim}\SI{2.5}{ns}$ bins to form the second-order correlation of photon arrival times, or $G^{(2)}(\tau)$ (after the aforementioned corrections are applied).

\begin{figure}[H]
    \centering
    \includegraphics[width=\columnwidth]{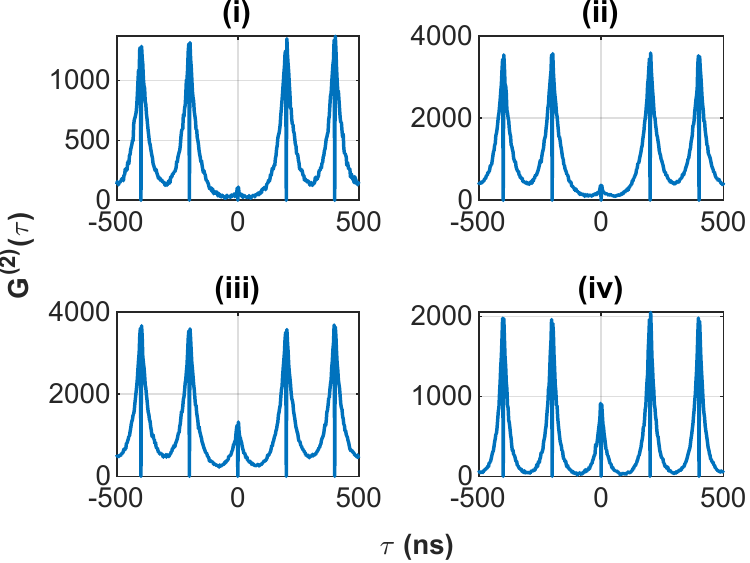}
    \caption{\textbf{Second-Order Correlation of Photon Arrival Times.} \(G^{(2)}(\tau)\) of the single particles shown in \autoref{fig:singleParticles} in the main text. (i) A monomer, a fused dimer with a (ii) high and a (iii) low \(g^{(2)}(0)\) contrast, and a (iv) non-fused dimer. The value of \(G^{(2)}(\tau)\) indicates the number of photon pairs detected $\tau$ apart. The ratio between the area under the central peak and the average area under the other peaks is termed the zero-delay normalized second-order correlation of photon arrival times ($g^{(2)}(0)$). To eliminate the dominant contribution of crosstalk at shorter $\tau$, the central bin of each peak is zeroed.}
    \label{SIfig:g2}
\end{figure}

\autoref{SIfig:g2} displays \(G^{(2)}(\tau)\) of the single particles shown in \autoref{fig:singleParticles} in the main text. It shows a series of peaks corresponding to the laser excitation period (\SI{200}{ns}). The ratio between the area under the central peak and the average area under the other peaks is termed the zero-delay normalized second-order correlation of photon arrival times ($g^{(2)}(0)$). To eliminate the dominant contribution of crosstalk at shorter $\tau$, the \(G^{(2)}(0)\) bin is zeroed. To avoid a biased area ratio between peaks with and without the exclusion of the peak point, the other peak points are also zeroed and excluded from the \(g^{(2)}(0)\) calculation. The particles shown display  \(g^{(2)}(0)\) values of (i) ${\sim}0.09$, (ii) ${\sim}0.13$, (iii) ${\sim}0.37$ and (iv) ${\sim}0.45$. 

\begin{figure}[H]
    \centering
    \includegraphics[width=\columnwidth]{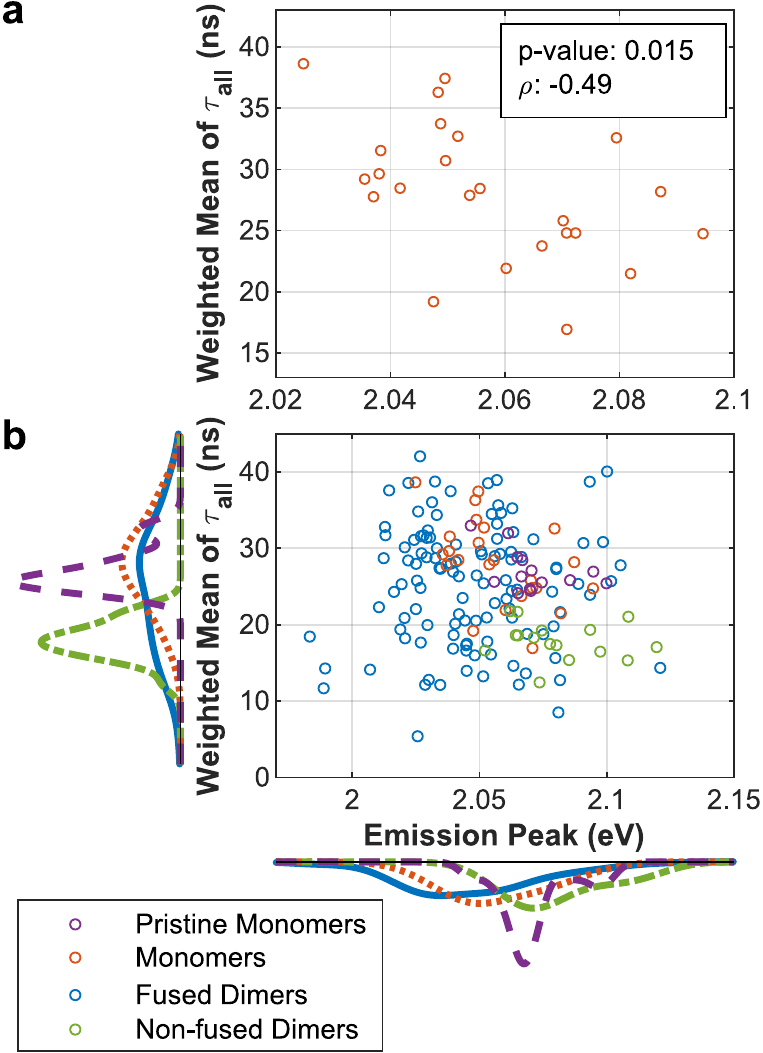}
    \caption{\textbf{Lifetime as a Function of the Peak of Emission Spectrum.} Weighted mean of the lifetimes of all detections as a function of the emission peak for \textbf{(a)} monomers only and for \textbf{(b)} all particles, colored according to particle type. p-value: p-value of Pearson's linear correlation. $\rho$: Pearson's linear correlation coefficient. In panel (b), lines to the left and beneath the axes represent the marginal distributions as kernel density plots, with colors matching the particle type.}
    \label{SIfig:specVsLT}
\end{figure}

\textbf{Comparison to Pristine Monomers.} \autoref{SIfig:specVsLT} shows the emission spectrum as a function of the overall lifetime ($\tau_{all}$), i.e., the weighted mean lifetime assessed for all detections, for each particle. All detections from a single measurement, binned according to the delay from their preceding laser pulse, are fitted to a bi-exponential decay model. The weighted mean of both lifetimes is then used to assess the overall lifetime. 
\autoref{SIfig:specVsLT}a displays this plot for monomers only and shows a negative correlation between the overall lifetime and emission energy. \autoref{SIfig:specVsLT}b shows the same plot, but for all particle types. The non-fused dimers and pristine monomers have a similar distribution in energy. This makes sense since pristine monomers are the building blocks of the non-fused dimers, both of which did not undergo fusion. Both particle types are slightly blue-shifted relative to monomers, that were exposed to the same fusion conditions as the fused dimers. This is due to the ripening process that occurs during fusion, which resulted in further shell growth and thus, a red-shift in emission energy. An additional red-shift is observed in fused dimers compared to monomers. We attribute this shift to the hybridization of the electron wave function and the additional shell volume in the neck region. 
 
\autoref{SIfig:specVsLT} also shows that the lifetimes of monomers and pristine monomers have similar distributions, while fused dimers exhibit slightly shorter lifetimes, possibly due to increased emission from charged and multi-excited states.\cite{Koley2022} Non-fused dimers display a significantly shorter decay lifetime than pristine monomers. This might be due to the architecture of the non-fused dimers, increasing charge-trapping in the region between the two QDs and increasing occupation of the charged exciton state, with a shorter lifetime than the neutral exciton state, thus shortening the effective decay lifetime of emission.

\begin{figure}[H]
    \includegraphics[width=\columnwidth]{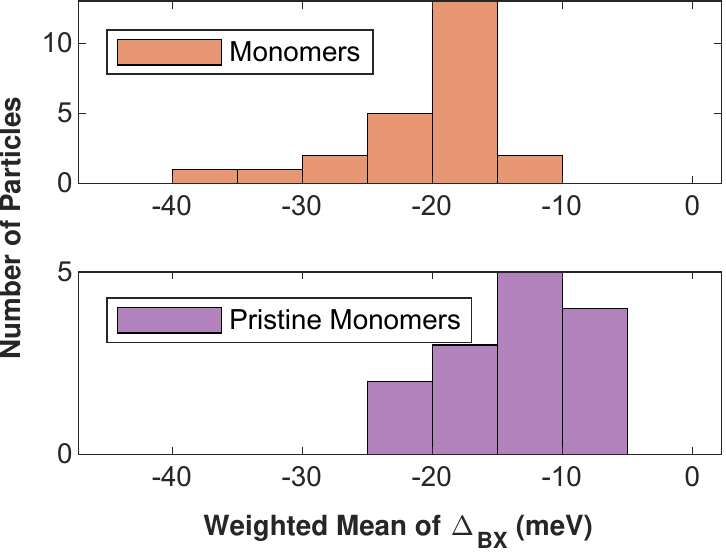}
    \caption{\textbf{BX Shifts of Monomers and Pristine Monomers.} The weighted mean of $\Delta_{BX}$ is calculated by averaging the BX shifts of the two BX components, weighting them according to their relative contribution.}
    \label{SIfig:BXshiftsMonomers}
\end{figure}

A further confirmation regarding the assumption of monomers’ shell growth is shown in \autoref{SIfig:BXshiftsMonomers}, which compares the weighted mean of the BX shifts ($\Delta_{BX}$) of the two BX components for monomers and pristine monomers. The weighted mean is calculated according to each component’s relative contribution, as shown in \autoref{fig:BXAgg} in the main text. Monomers exhibit a slightly stronger BX shift than pristine monomers. This stronger shift agrees with the suggested shell growth in monomers, allowing electrons to delocalize further into the shell. Thus, hole--hole repulsion is given more weight, leading to further blue-shift.\cite{Oron2007}

Additionally, we note that the weighted mean of $\Delta_{BX}$ for pristine monomers (in \autoref{SIfig:BXshiftsMonomers}) resembles the $\Delta_{BX}$ of the fast BX component of non-fused dimers (bottom panel of \autoref{fig:BXHist}a in the main text). The fast BX component represents the LBX, which is comparable to BX states in monomers. Accordingly, we see similarities in the fast BX component's $\Delta_{BX}$ of monomers and fused dimers, which underwent fusion, and of pristine monomers and non-fused dimers, which did not undergo fusion.

\textbf{$g^{(2)}(0)$ of Two Uncorrelated Photo-emitters and the Non-fused Dimers Sample.} At the low excitation regime of this experiment ($\langle{N} \rangle{\ll}1$), the photon antibunching is expected to be proportional to the BX quantum yield (QY) so that $g^{(2)}(0){\sim}\frac{QY_{BX}}{QY_{1X}}$.\cite{Nair2011} If we assume unity for the $QY_{1X}$, then $g^{(2)}(0){\simeq}QY_{BX}$. To set an upper bound for the expected $g^{(2)}(0)$ for non-fused dimers, we consider them as two uncorrelated monomers. To be more accurate, pristine monomers are the building blocks of non-fused dimers, featuring a $g^{(2)}(0)$ distribution of $0.09{\pm}0.02$, which is similar to monomers ($g^{(2)}(0){=}0.1{\pm}0.03$). The calculation here is done by counting the probabilities for the different configurations of two detections following the same laser pulse ($G^{(2)}(\tau)$ central peak, or $G^{(2)}(center)$; see \autoref{SIfig:g2}) and sequential laser pulses ($G^{(2)}(\tau)$ side peaks, or $G^{(2)}(\infty)$). 
Considering two uncorrelated monomers, termed as emitter “A” and emitter “B”, the possible configurations that contribute to $G^{(2)}(\infty)$ are A-A, B-B, A-B, and B-A, corresponding to the probability of two emissions a laser period ($\SI{200}{ns}$) apart. Each character represents an emission from the corresponding emitter ("A" or "B"). Since we assume $QY_{1X}{\approx}1$, this sums up to $G^{(2)}(\infty){=}4$. The possible configurations that contribute to $G^{(2)}(center)$ are AA, BB, AB, and BA, corresponding to the probability of two emissions following a single excitation pulse. The AA and BB configurations represent the probability of a BX--1X emission cascade in a monomer. Hence, they are equal to $QY_{BX, monomer}{\simeq}g^{(2)}(0){=}0.1{\pm}0.03$. In contrast, the AB and BA configurations represent two non-interacting single excitons. Hence their probability is 1, as shown before. Consequently, the expected $g^{(2)}(0)$ contrast for two uncorrelated single photon emitters is 
$g^{(2)}(0)= \frac{G^{(2)}(center)}{G^{(2)}(\infty)}{\approx} \frac{2\cdot1+ 2\cdot0.1}{4\cdot1}{=}0.55$.

\begin{figure}[H]
    \includegraphics[width=\columnwidth]{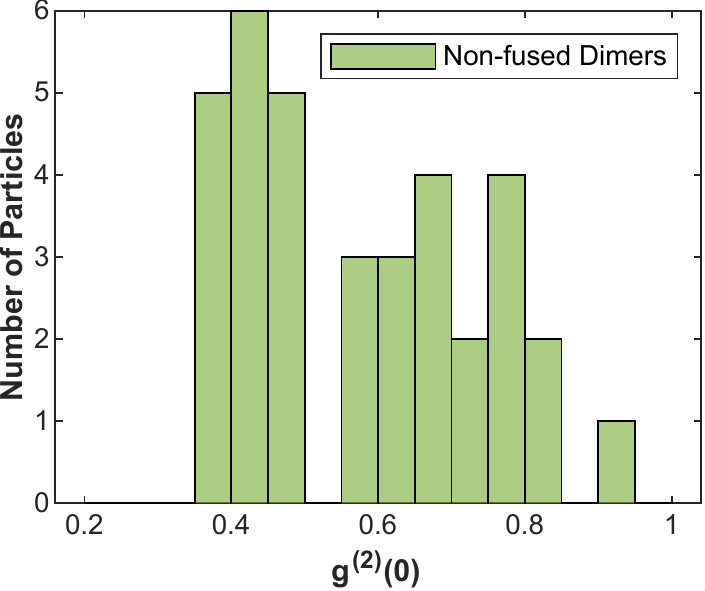}
    \caption{\textbf{$g^{(2)}(0)$ Values from the Non-fused Dimers Sample.} Photon antibunching for all particles from the non-fused dimers sample (except ones classified as monomers), without filtering for $g^{(2)}(0){<}0.55$.}
    \label{SIfig:nonfusedG2}
\end{figure}

We note that the non-fused dimers measured can be categorized into two distinct populations, with $g^{(0)}(2){<}0.55$ and $g^{(0)}(2){>}0.55$ (see \autoref{SIfig:nonfusedG2}). The first population suggests that even for the non-fused dimers, some interaction between segregated excitons may exist, which leads to enhanced antibunching. We attribute the second category of $g^{(0)}(2){>}0.55$ to contribution from oligomers or charged states. The architecture of the non-fused dimers may increase charge trapping in the region between the two QDs. The charged excitons will undergo fast Auger decay and decrease the 1X and BX QYs. As the QY of the 1X is much more sensitive to charging than that of the BX,\cite{Xu2017} $g^{(2)}(0)$ contrasts might exceed expected values, for $g^{(2)}(0){\sim}\frac{QY_{BX}}{QY_{1X}}$ at the $\langle{N}\rangle\ll{1}$ regime of this experiment (\autoref{SIfig:saturation}).\cite{Nair2011}  

\autoref{SIfig:nonfusedG2} displays all particles from the non-fused dimers sample (except ones classified as monomers). Two distinct populations can be observed: one with $g^{(2)}(0){<}0.5$ and another with higher values. To avoid the possible inclusion of oligomers or highly charged particles, in this work we set a threshold of $g^{(2)}(0){<}0.55$, and omitted non-fused dimers with higher $g^{(2)}(0)$ values.

 \begin{figure}[H]
    \centering
    \includegraphics[width = \columnwidth]{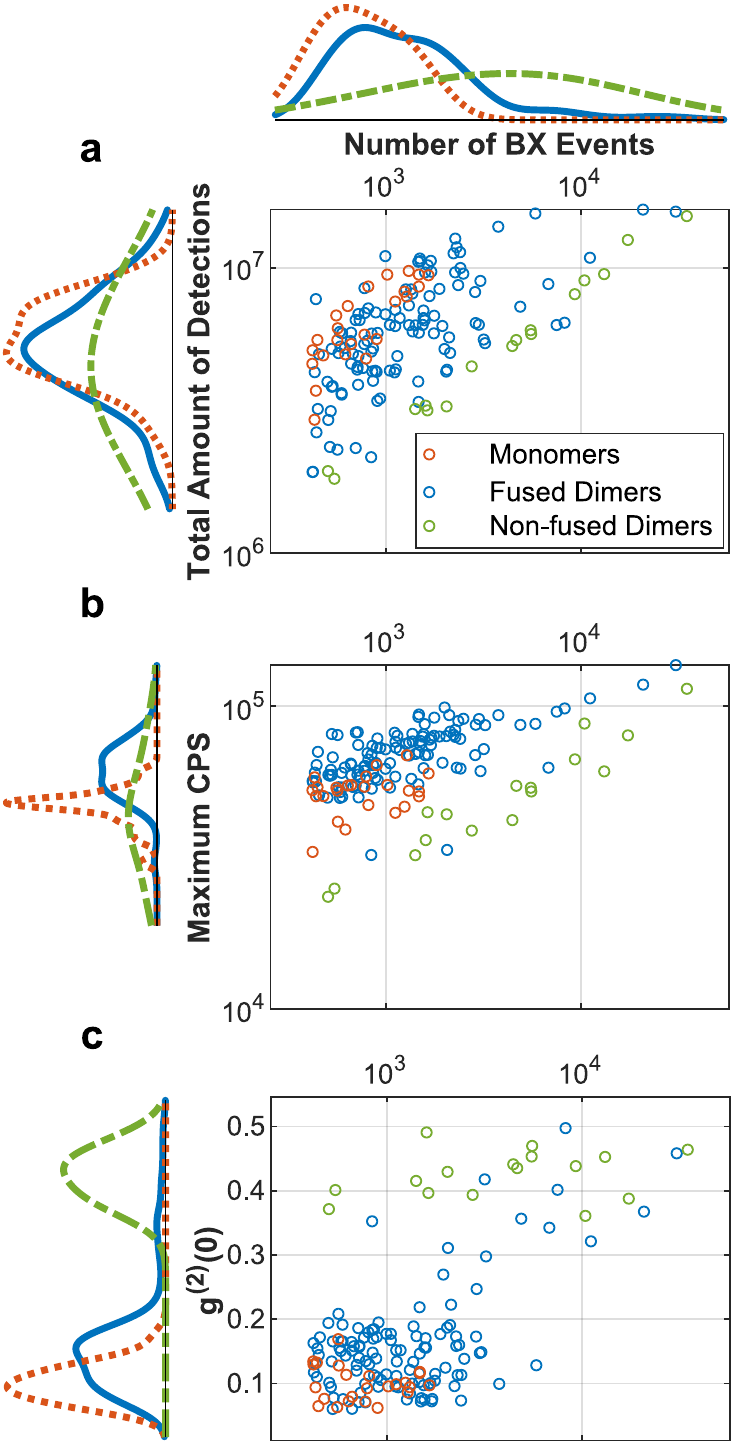}
    \caption{\textbf{Number of Heralded Events, Brightness and $g^{(2)}(0)$.} The number of BX events is shown as a function of \textbf{(a)} the total amount of detections in a 5-minute measurement, \textbf{(b)} the maximal counts per second (CPS) in a \SI{10}{ms} bin, and \textbf{(c)} the $g^{(2)}(0)$, colored according to type. Lines to the left and above the axes represent the marginal distributions as kernel density plots, with colors matching the particle type.}
    \label{SIfig:BXNum}
\end{figure}

\textbf{BX Detection Count.} \autoref{SIfig:BXNum}a features the number of BX detection events as a function of the total amount of detections for each 5-minute measurement. The excitation probability of single or multiple excitons follows the Poisson distribution.\cite{Peterson2009} Hence, for bright single particles, dominated by radiative exciton emission, more BX emission is expected as well. Therefore, a positive correlation is observed between the two variables. \autoref{SIfig:BXNum}a shows that on average, fused dimers are about as bright as monomers, whereas \autoref{SIfig:BXNum}b shows that upon binning the detections ($\SI{10}{ms}$ bins), fused dimers exhibit higher maximal brightness (the bin with the most detections is taken as the maximum momentary brightness). These two observations support prior claims regarding the prevalence of charged states in dimers.\cite{Koley2022} Fused dimers have a higher absorption cross-section than monomers and therefore present a brighter `on’ state (as appears in \autoref{SIfig:BXNum}b).\cite{Panfil2019} However, they are frequently charged, which reduces their overall counts.

As mentioned earlier, at the low excitation regime of this experiment ($\langle{N}\rangle{\ll}1$), the photon antibunching ($g^{(2)}(0)$) is expected to be proportional to the BX quantum yield.\cite{Nair2011} Accordingly, a strong correlation between the $g^{(2)}(0)$ and the number of BX events is observed in \autoref{SIfig:BXNum}c.

\begin{figure}[H]
    \centering  
    \includegraphics[width=\columnwidth]{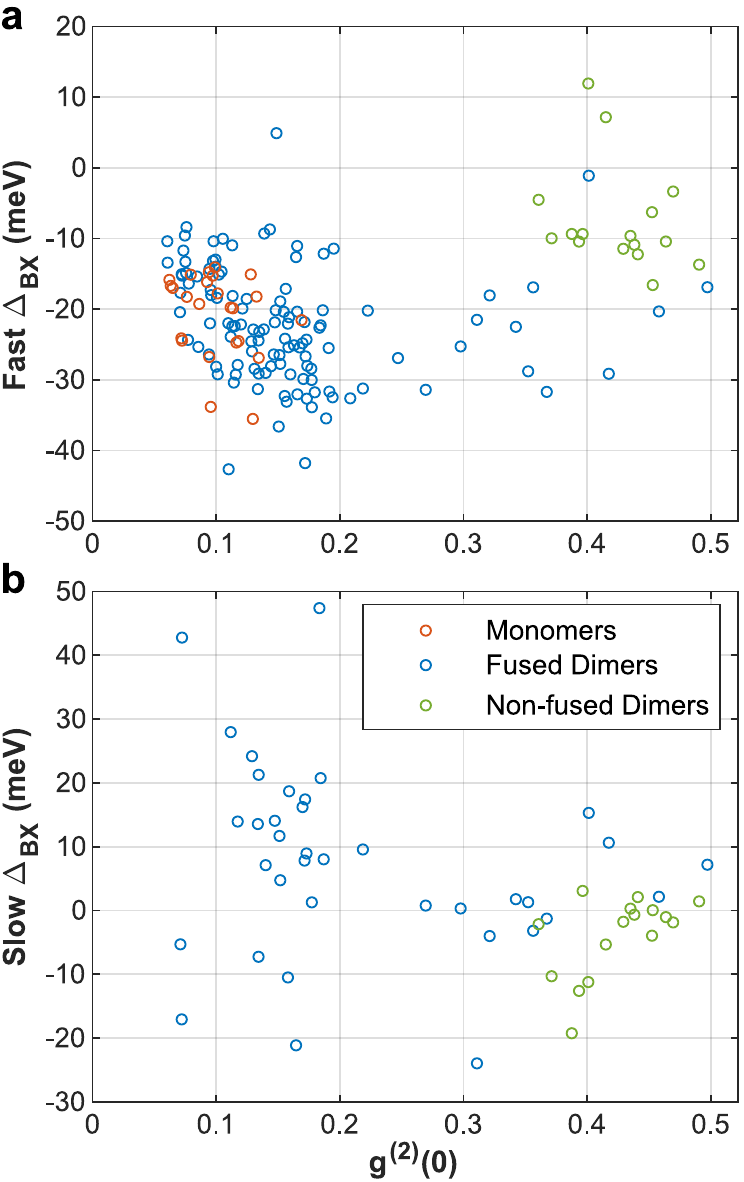}
    \caption{\textbf{BX Shifts as a Function of $g^{(2)}(0)$.} BX shifts ($\Delta_{BX}$) of the \textbf{(a)} fast and the \textbf{(b)} slow BX components of each particle, as a function of photon antibunching. Most of the fused dimers and monomers do not appear in (b) because they do not exhibit a slow BX component at all.}
    \label{SIfig:BXshifts}
\end{figure}

\textbf{BX Components Shift.} \autoref{fig:BXAgg} in the main text displays a variation in the mean BX shift and BX lifetime as a function of $g^{(2)}(0)$, especially for fused dimers. The fused dimers' distribution of the fast BX shift is uncorrelated with variation in $g^{(2)}(0)$ (\autoref{SIfig:BXshifts}a). Non-fused dimers exhibit a slightly weaker fast BX shift, however, this is attributed to the increasing confinement in their constituent thinner-shell QDs, as explained in the main text. In contrast to the fast component, the BX shift of the slow BX component in fused dimers does show a stronger correlation with $g^{(2)}(0)$ (\autoref{SIfig:BXshifts}b). Nevertheless, it shows a negative correlation with $g^{(2)}(0)$, opposite from the trend observed in \autoref{fig:BXAgg} in the main text. The slow BX component contribution, however, is positively correlated with $g^{(2)}(0)$ (\autoref{fig:BXAgg}c). This trend corresponds to our suggestion in the main text that the variations in the mean BX shift and BX lifetime are carried out by the changing ratio between the contributions of the two BX components, and not by a change in the BX states' energetics.    

\begin{figure}[H]
    \centering
    \includegraphics[width=\columnwidth]{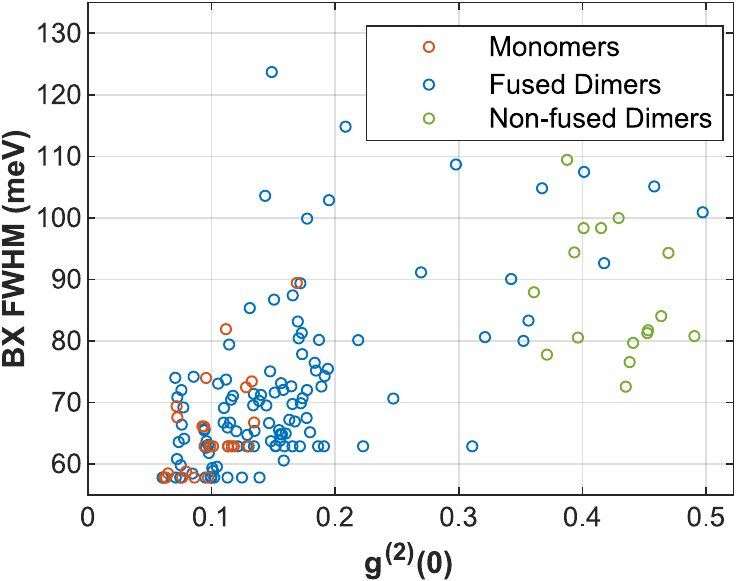}
    \caption{\textbf{BX Spectral Width.} The full width at half maximum (FWHM) of the whole BX emission, according to its fit to a Voigt profile, as a function of $g^{(2)}(0)$, colored according to particle type.}
    \label{SIfig:BXFwhm}
\end{figure}

\textbf{BX Spectral Width.} The overall BX spectral width is assessed here by fitting the whole BX population (without separation into different components) to a Voigt profile. In \autoref{SIfig:BXFwhm}, the full width at half maximum (FWHM) for each particle, according to type, is plotted against $g^{(2)}(0)$. The BX FWHM increases with $g^{(2)}(0)$, which is correlated with an increase in SBX emission over LBX, as observed in the increasing contribution of the slow BX component in \autoref{fig:BXAgg}c in the main text. This agrees with the expected BX spectral broadening due to an increased emission by transitions 4 and 5 (see \autoref{fig:simulation} in the main text) as the SBX emission increases. This is compared to a spectrally narrow BX emission expected for particles with a dominant LBX emission (monomers and most of the fused dimers), that would emit BXs mainly through transition 3. 

\begin{figure}[H]
    \centering
    \includegraphics[width=\columnwidth]{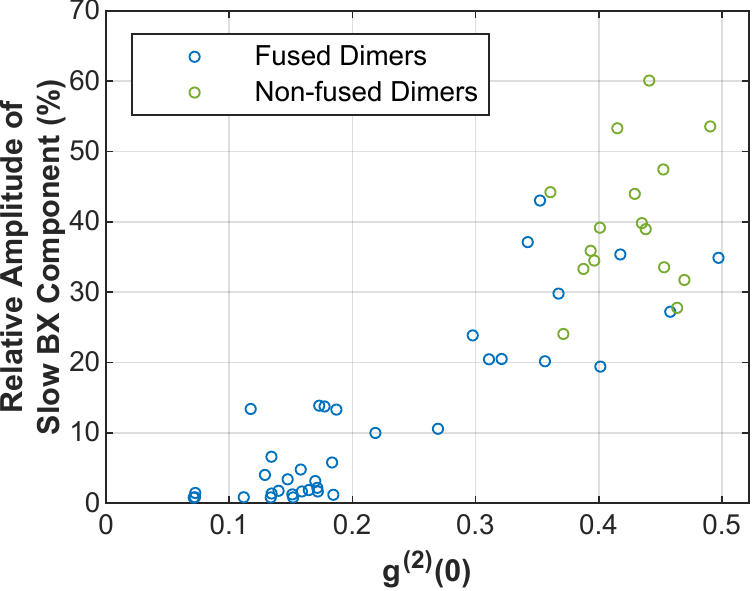}
    \caption{\textbf{Slow BX Component Initial Occupation.} The relative amplitude $\frac{\frac{a_1}{\tau_1}}{\sum_{i=1}^{2} \frac{a_i}{\tau_i}}$ of the slow BX component is shown as a function of $g^{(2)}(0)$. Only particles that featured some contribution of the slow (lifetime of ${>}\SI{1}{ns}$) BX component (as defined in the main text) are shown.}
    \label{SIfig:slowBXAmp}
\end{figure}

\textbf{BX Initial Occupation.} To study the initial occupation of the different BX species, we examine the relative amplitude of the two components, i.e., $\frac{\frac{a_i}{\tau_i}}{\sum_{i=1}^{2} \frac{a_i}{\tau_i}}$ from \autoref{eq:model} in the main text. This quantity is attributed to the relative weight of each component at $t=0$ (compared to $\frac{a_i}{\sum_{i=1}^{2} a_i}$ which means the total contribution of each component between $t=0$ and $t=\infty$; see \autoref{fig:BXAgg}c in the main text). Following the Poisson distribution, the probabilities to initiate each of the BX states is $P_{LBX}{\simeq}2P_{BX}{=}2(1-(1+ \langle N \rangle) e^{-\langle N \rangle})$ and $P_{SBX}{\simeq}P_{1X}^2{=}(1-e^{-\langle N \rangle})^2$. In the $\langle N \rangle \ll 1$ regime of this experiment, their ratio is ${\sim} 1$, so we would expect that the initial excitations of the LBX and of the SBX states would be with a similar probability. This agrees well with the similar relative amplitude for the fast and slow BX components (which we attribute to the LBX and the SBX, respectively), observed in \autoref{SIfig:slowBXAmp} for non-fused dimers (relative amplitude of $40{\pm}10\%$ for the slow component). 

In comparison, all fused dimers exhibit a greater amplitude for the fast BX component (slow component relative amplitude ${<}50\%$). This agrees with our assumption of exciton kinetics coming into play in our observations. As the neck size increases (i.e., $g^{(2)}(0)$ values decrease), we assume that the inter-dot transfer mechanisms become much faster than the radiative lifetime ($\tau_T{\ll}\tau_r$). Therefore, the transition of SBX to LBX (which we suggest is the BX ground state) is faster than our detection resolution,  resulting in an apparent increased occupation of the LBX state over the SBX. In addition, for a homodimer case, the SBX state would be the ground state, which will result in a greater SBX initial occupation. All of the fused dimers and most of the non-fused dimers have a relative slow BX amplitude of ${<}50\%$, showing the dominance of the LBX state, which also supports our assumption regarding the heterogeneity of all dimers.

\section{Quantum Mechanical Simulations}

\begin{figure}[H]
    \centering
    \includegraphics[width=\columnwidth]{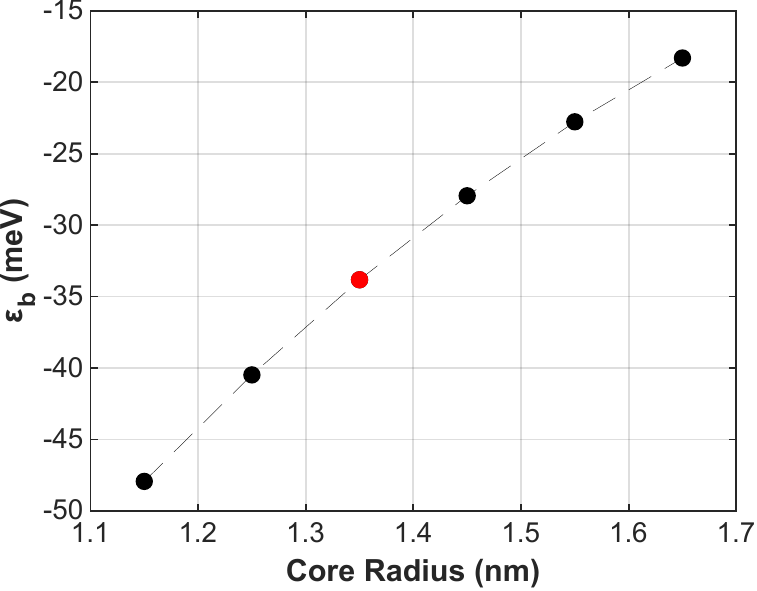}
    \caption{\textbf{Calculated Monomer BX Binding Energy.} The BX binding energy of a monomer with a fixed total radius (core + shell) of $\SI{3.4}{nm}$ as a function of the core radius. The black dots are calculated values, while the black dashed line is a guide to the eye. The red dot indicates the approximate core size of the particles studied experimentally.}
    \label{SIfig:monomerEbSimulation}
\end{figure}

\textbf{Calculated Monomer BX Binding Energy.} \autoref{SIfig:monomerEbSimulation} features the calculated BX binding energy of a monomer as a function of its core radius. The total core and shell radius is fixed to $\SI{3.4}{nm}$ while the core radius alone is varied. For the core radius in this experiment (${\sim}\SI{1.35}{nm}$) $\varepsilon_b{\approx}-\SI{34}{meV}$, which is a stronger 1X--1X interaction than the one observed experimentally (see \autoref{fig:BXHist}a in the main text).

\begin{figure}[H]
    \centering
    \includegraphics[width=\columnwidth]{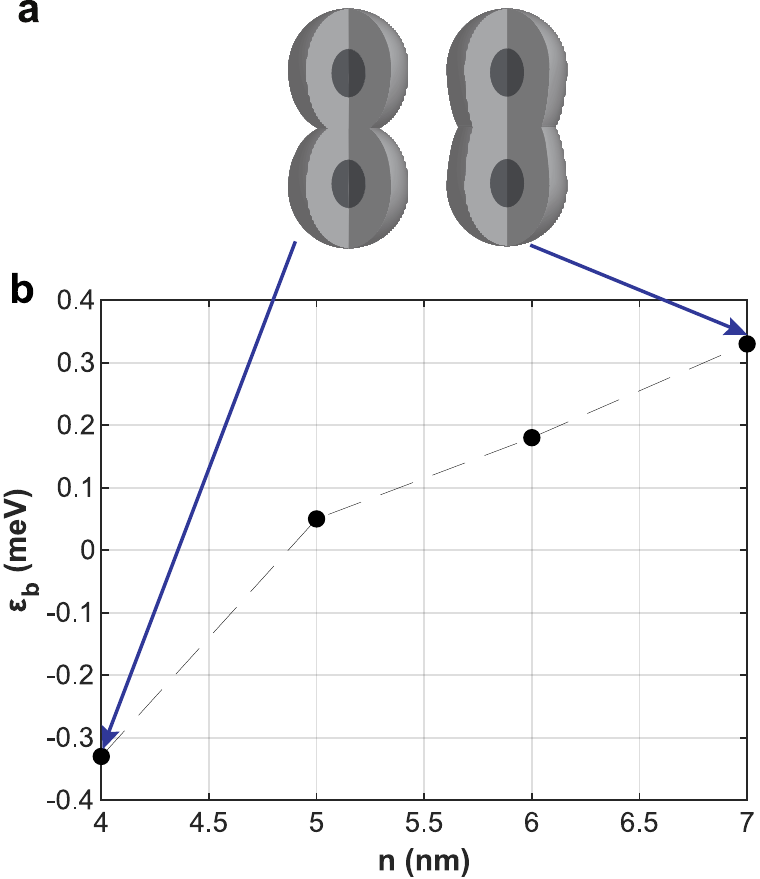}
    \caption{\textbf{Neck Size Effect on BX Binding Energy.} \textbf{a)} Illustrations of homodimers with a small (left) and a large (right) neck size. \textbf{b)} The BX binding energy of a homodimer is shown as a function of $n$, which is the semi-axis of the ellipsoidal shell in the coupling direction of the dimer. An increase in $n$ corresponds to a larger neck size, as explained in the main text. The black dots are calculated values, while the black dashed line is a guide to the eye.}
    \label{SIfig:simulatedNeckEffect}
\end{figure}

\textbf{Neck Size Effect on BX Binding Energy.} \autoref{SIfig:simulatedNeckEffect} exhibits the calculated effect the neck size has on the BX binding energy in homodimers. The BX binding energy is always close to zero, which is a result of the weak inter-dot 1X--1X interaction in the SBX state (i.e., the lower-energy BX state in homodimers; see main text). The neck size has a minor influence on the electronic structure because the wave functions of both the electrons and holes are mostly localized in or around the respective cores. The neck was found to have a negligible effect on the BX binding energy in the heterodimer case as well. Notice that the y-axis units in \autoref{SIfig:simulatedNeckEffect}b are in meVs, displaying a change in $\varepsilon_b$ of ${<}\SI{1}{meV}$ while varying from a dimer with almost no connecting neck ($n{=}\SI{4}{nm}$) to a one with a `rod-like' geometry ($n{=}\SI{7}{nm}$).

\bibliographystyle{achemso}
\bibliography{supp.bib}
\end{multicols}

\makeatletter\@input{yy.tex}\makeatother